\documentclass[11pt, oneside]{article} 
\usepackage[utf8]{inputenc}
\usepackage[utf8]{inputenc}
\usepackage{geometry}                		
\usepackage{graphicx}				
\usepackage{amssymb}
\usepackage{amsmath}
\usepackage{subfig}
\usepackage{caption}
\usepackage{comment}
\usepackage{amsthm,verbatim,amsfonts,amscd,float,physics}
\usepackage{appendix}

\usepackage[utf8]{inputenc}
\usepackage{xcolor}
\numberwithin{equation}{section}
\usepackage{titling, lipsum}
\title{Quantum Computing for \\
Rotating, Charged and String Theory Black Holes}
\author{ Viti Chandra$^{(1)}$, Michael McGuigan$^{(2)}$\\
(1) Half Hollow Hills High School West \\
(2) email contact: michael.d.mcguigan@gmail.com
}
\date{}

\begin{document}
\begin{titlingpage}

\maketitle
\begin{abstract}
The quantum mechanics of Rotating, Charged, de Sitter and String Theory black holes are of recent interest because of their peculiar thermodynamic properties, as well the mysterious nature of their microstates. A full quantum treatment of the operators involved in this systems could yield valuable information into their nature, similar to how quantum treatment yields valuable insight into atoms, molecules and elementary particles. We study four types of black holes using quantum computing, which include the 3D Rotating Banados-Teitelboim-Zanelli (BTZ) black hole, the 4D charged Reisner-Nordtrom (RN) black hole, the 4D charged Reisner-Nordstrom -de Sitter (RN-dS) black hole and the 2D charged string black hole. In these cases in addition to the Hamiltonian there is a Mass operator which plays an important role in describing the quantum states of the black hole. We compute the spectrum of these operators using classical and quantum computing. For quantum computing we use the Variational Quantum Eigensolver (VQE) which is hybrid classical-quantum algorithm that runs on near term quantum hardware. We perform our calculations using 4 qubits in both a harmonic oscillator and position basis, realizing the quantum operators of the black holes  in terms of $16\times 16$  matrices. For the 4 qubit case we find highly accurate results for the Mass eigenvalues for different values of the charge and angular momentum. For the 2D Charged String black hole we also use the VQE to compute the expectation value of the Hamiltonian constraint and the commutator of the Hamiltonian constraint with the mass operator and find excellent agreement with theoretical expectations.

\end{abstract}
\end{titlingpage}
\section{Introduction}

The quantum treatment of black holes is of great interest. There are mysterious quantum aspects of black hole entropy and thermodynamics. Additionally, there are quantum effects in the black hole interior where the curvature is so large that quantum gravity becomes important. Quantum computing represents a potentially disruptive computing paradigm for the acceleration of certain applications. One such application is quantum simulation, where the simulations are implemented on a quantum computer by representing the Hamiltonian and states as a matrix with qubits and expanding the matrix in terms of Pauli tensor products. One then implements the states and Unitary operations in therms of products of a basic set of gates which implement quantum operations like superposition and entanglement. The quantum circuit is evaluated using a measurement operation on the final state of the right side of the circuit. Usually these quantum simulations are applied to quantum chemistry or materials science. There have been recent attempts to expand the range of applications of quantum computing to quantum cosmology, matrix models and black holes where quantum gravity plays a role \cite{Joseph:2022ggp}
\cite{Joseph:2021naq}
\cite{Kaufman:2019vzn}
\cite{Kocher:2018ilr}
\cite{Ganguly:2019kkm}
\cite{bhqc1}
\cite{bhqc2}
\cite{Antonini:2019qkt}
\cite{Brown:2019hmk}
\cite{Nezami:2021yaq}
\cite{Rinaldi:2021jbg}.
In particular in \cite{Joseph:2022ggp} quantum computing was applied the Schwarzschild and Schwarzschild-de Sitter black holes obtaining very accurate results for lowest mass quantum states. In this paper we extend that analysis to charged and rotating black holes. We examine black holes with different characteristics in different space time dimensions and compare the accuracy of the quantum computations in different bases used to describe the quantum states. Table 1 gives the basic characteristics of the black holes we study which include the Banados-Teitelboim-Zanelli (BTZ) rotating black hole in three spacetime dimensions, the Reisner-Nordtrom black hole and Reisner-Nordstrom -de Sitter black hole in four dimensions and the charged string black hole in two space-time dimensions. 

This paper is organized as follows. In section two we describe the basic equations, metric, entropy and temperature of the four types of black hole we consider. In section three we discuss the basic quantum computing  methods and quantum computing algorithms we apply and how they are implemented for the four types of black holes. In section four we present out quantum computing results and discuss the accuracy of the methods for each of the black holes with regards to the classical computations. In section five we give the main conclusions of the paper.

\begin{table}[ht]
\centering
\begin{tabular}{|l|l|l|l|l|}
\hline
Black hole type & space-time dimensions & Charged & Rotating & Cosmological constant   \\ \hline
BTZ black hole & 3 & N & Y  & negative  \\ \hline
RN black hole & 4 & Y & N & zero \\ \hline
RN-dS black hole & 4 & Y & N & positive \\ \hline
String black hole & 2 & Y & N & negative \\ \hline

\end{tabular}
\caption{\label{tab:BasisCompare}  Characteristics of the four types of black holes that we apply quantum computing techniques in this paper.}
\end{table}

\section{Rotating and charged black holes}

In this section we give a description of the four types of black holes that that we will will consider using quantum computing.

\subsection*{3D rotating BTZ black hole}

Perhaps the simplest example of a rotating black hole is the BTZ solution in three space-time dimensions \cite{Banados:1992wn}
\cite{Carlip:1995qv}. The action for this type of black hole is given by:
\[{S_{3d}} = \frac{1}{{16\pi G }}\int {{d^3}x\sqrt { - g} } \left( {R - 2\lambda } \right)\]
As we take $\lambda$ negative in three space-time dimensions we set  $\lambda  =  - \frac{1}{{{\ell ^2}}}$ and we choose units so that $G=\frac{1}{8}$.
Using the ansatz:
\[d{s^2} =  - {N^2}d{t^2} + a^2d{r^2} + {b^2}{\left( {d\varphi  + {N^\varphi }dt} \right)^2}\]
where $N,a,b, N^\varphi$ are functions of $r$, the BTZ black hole solution is
\begin{align}
&N = {\left( { - M + \frac{{{r^2}}}{{{\ell ^2}}} + \frac{{{J^2}}}{{4{r^2}}}} \right)^{1/2}}\nonumber\\
&a = {\left( { - M + \frac{{{r^2}}}{{{\ell ^2}}} + \frac{{{J^2}}}{{4{r^2}}}} \right)^{ - 1/2}}\nonumber\\
&b = r\nonumber \\
&{N^\varphi } =  - \frac{{J}}{{{2r^2}}}
\end{align}

Here $M, J$ are parameters related to the mass and angular momentum of the black hole. To describe the rotating black hole, it is useful to have expressions for the Mass and Angular momentum of the black hole system in terms of $N,a,b,N^{\varphi}$. These were derived in \cite{Achucarro:1993fd}
 using the first order formalism and are given by:
\begin{align}
&M = \omega _\varphi ^a{\omega _{a\varphi }} + \frac{1}{{{\ell ^2}}}e_\varphi ^a{e_{a\varphi }}\nonumber \\
&J = 2e_\varphi ^a{\omega _{a\varphi }}\\
\end{align}
Where
\begin{align}
&{e^{(0)}} = Ndt\nonumber \\
&{e^{(1)}} = adr\nonumber \\
&{e^{(2)}} = b({N^\varphi }dt + d\varphi )\\
\end{align}
and
\begin{align}
&{\omega ^{(0)}} = \left( { - \frac{1}{2}bN{{\left( {{N^\varphi }} \right)}^\prime } - b'N{N^\varphi }} \right)dt - b'Nd\varphi \nonumber \\
&{\omega ^{(1)}} =  - \frac{1}{{2N}}b{\left( {{N^\varphi }} \right)^\prime }dr\nonumber \\
&{\omega ^{(2)}} = \left( {N{{\left( N \right)}^\prime } + \frac{1}{2}{b^2}{N^\varphi }{{\left( {{N^\varphi }} \right)}^\prime }} \right)dt + \frac{1}{2}{b^2}{\left( {{N^\varphi }} \right)^\prime }d\varphi 
\end{align}
with M the mass and J the angular momentum than become
\begin{align}
&M = \frac{{{{\dot b}^2}}}{{{N^2}}} - \frac{{{{b'}^2}}}{{{a^2}}} + \frac{{{b^2}}}{{{\ell ^2}}} + \frac{{{J^2}}}{{4{b^2}}}\nonumber \\
&J = {b^3}{N^\varphi }^\prime 
\end{align}
Note that the canonical momentum associated with the variable $a$ is $\frac{\dot b}{N}$ so the mass operator can be written:
\begin{equation}M = p_a^2 - \frac{{{{b'}^2}}}{{{a^2}}} + \frac{{{b^2}}}{{{\ell ^2}}} + \frac{{{J^2}}}{{4{b^2}}}\end{equation}
The outer  and inner horizon of the BTZ black hole are:
\begin{equation}{r_ \pm } = \frac{\ell }{2}\left( {\sqrt {M + \frac{J}{\ell }}  \pm \sqrt {M - \frac{J}{\ell }} } \right)\end{equation}
The entropy of the rotating black hole is proportional to $r_+$ and is plotted in figure 
\begin{equation}r_ \pm ^2 = \frac{{{\ell ^2}}}{2}\left( {M \pm \sqrt {{M^2} - \frac{{{J^2}}}{{{\ell ^2}}}} } \right)\end{equation}
Then the entropy, inverse temperature and temperature are given by:
\begin{align}
&S = \frac{{2\pi {r_ + }}}{{4G}}\nonumber \\
&\beta  = \frac{{2\pi {r_ + }{\ell ^2}}}{{r_ + ^2 - r_ - ^2}}\nonumber \\
&T = \frac{{r_ + ^2 - r_ - ^2}}{{2\pi {r_ + }{\ell ^2}}}
\end{align}
These expressions resemble the entropy density of a 1+1 dimensional quantum field theory which is evidence for the $AdS_3/CFT_2$ holographic duality associated with the BTZ black hole.

\subsection*{4D Charged RN Black Holes}

For charged black holes we study the Reisner-Nordstrom solution in de Sitter space with positive cosmological constant. Here the action is given by:
\begin{equation}S_{4d}=\frac{1}{16 \pi G}\int {{d^4}x\sqrt { - g} } \left( {R   - 2\lambda  - \frac{G}{4}{F_{\mu \nu }}{F^{\mu \nu }}} \right)\end{equation}
The metric is given by:
\begin{equation}d{s^2} =  - {N^2}d{t^2} + {a^2}d{r^2} + {b^2}d\Omega _2^2\end{equation}
with:
\begin{align}
&N = {\left( {1 - \frac{{2M}}{r} + \frac{{{Q^2}}}{{{r^2}}} - \frac{{\lambda {r^2}}}{3}} \right)^{1/2}}\nonumber \\
&a = {\left( {1 - \frac{{2M}}{r} + \frac{{{Q^2}}}{{{r^2}}} - \frac{{\lambda {r^2}}}{3}} \right)^{ - 1/2}}\nonumber \\
&b = r
\end{align}
the outer and inner horizons are located at:
\begin{equation}{r_ \pm } = M \pm \sqrt {{M^2} - {Q^2}} \end{equation}
and the entropy, inverse temperature and temperature are given by:
\begin{align}
&S = \frac{{4\pi r_ + ^2}}{{4G}} = \pi {\left( {M + \sqrt {{M^2} - {Q^2}} } \right)^2}\nonumber \\
&\beta  = \frac{{2\pi {{\left( {M + \sqrt {{M^2} - {Q^2}} } \right)}^2}}}{{\sqrt {{M^2} - {Q^2}} }}\nonumber \\
&T = \frac{{\sqrt {{M^2} - {Q^2}} }}{{2\pi {{\left( {M + \sqrt {{M^2} - {Q^2}} } \right)}^2}}}
\end{align}

\subsection*{4D Charged RN-dS black hole}

For the de Sitter charged black hole we set $\lambda=\frac{3}{\ell^2}$. 
The Hamiltonian, Momentum and Gauss law constraints are given by \cite{Kiefer:1998rr}:
\begin{align}
&H =  - \frac{{{p_a}{p_b}}}{b} + \frac{{a(p_a^2 + p_{A_1}^2)}}{{2{b^2}}} + \frac{{bb''}}{a} - \frac{{b b' a'}}{{{a^2}}} + \frac{{{{b'}^2}}}{{2a}} - \frac{a}{2} + \frac{{3a{b^2}}}{{2{\ell ^2}}}\nonumber \\
&{H_r} =  - a'{p_a} + {p_b}b' - A_1 {{p}_{A_1}'} \nonumber \\
&G =  - {{p}_{A_1}'}
\end{align}
The expressions for the Mass and Charge are given by:
\begin{align}
&M = \frac{{Gp_a^2}}{{2b}} + \frac{b}{{2G}}\left\{ {1 - \frac{{b'b'}}{{{a^2}}} - \frac{{\lambda {b^2}}}{3}} \right\} + \frac{Q^2}{2b}\nonumber \\
& Q = -b^2 A_0'
\end{align}
To quantize these expressions with quantum computing, we will represent $a,p_a,b,p_b$ in terms of $2^q \times 2^q$ matrices with $q$ the number of qubits, which we discuss in the next section.



\subsection*{2D charged string black hole}

Black holes in string theory involve the dilaton field
\cite{Garfinkle:1990qj}\cite{Gregory:1992kr}
\cite{Horne:1992bi}
\cite{Horne:1992zy}
. In a simpler context, in 1+1 dimensions there are analytic solutions for neutral and charged black holes with the dilaton field 
\cite{Witten:1991yr}
\cite{McGuigan:1991qp}
\cite{Nappi:1992as}
\cite{Kumar:1994ve}
\cite{Kumar:1994ce}
\cite{Giveon:1993hm}
\cite{Sfetsos:1993bh}
\cite{Suresh:2015bsa}. These can be used as toy models to study black holes in string theory. The dilaton field plays an important role in the 2D black hole  with the value of the dilaton field at the horizon effectively playing the role of the horizon radius. The 2D heterotic effective sting action of gravity, dilaton and gauge field is 
\begin{equation}S_{2d} =\int {{d^2}x\sqrt { - g} } {e^{ - 2\phi }}\left( {R + 4{{\left( {\nabla \phi } \right)}^2} - 2\lambda  - \frac{1}{4}{F_{\mu \nu }}{F^{\mu \nu }}} \right)\end{equation}
To obtain black hole solutions $\lambda$ must be negative and we set $\lambda = - \frac{1}{2\ell^2}$
For the Heterotic charged black hole  we use the ansatz:
\begin{equation}d{s^2} =  - {N^2}d{t^2} + {a^2}d{r^2}\end{equation}
and the black hole solution is given by:
\begin{align}
&N = {\left( {1 - 2m{e^{ - r/\ell }} + {q^2}{e^{ - 2r/\ell }}} \right)^{1/2}}\nonumber \\
&a = {\left( {1 - 2m{e^{ - r/\ell }} + {q^2}{e^{ - 2r/\ell }}} \right)^{ - 1/2}}\nonumber \\
&b = {e^{ - \phi }} = {e^{ - {\phi _0} + r/2\ell }}
\end{align}
and we have defined $b=e^{-\phi}$. The Electric field of the 2D charged heterotic black hole is:
\begin{equation}{F_{tr}} = \frac{{\sqrt 2 }}{\ell }q{e^{ - r/\ell }}\end{equation}
The horizons are located at:
\begin{equation}{r_ \pm } = \ell \log \left( {m \pm \sqrt {{m^2} - {q^2}} } \right)\end{equation}
The entropy, inverse temperature and temperature are given by:
\begin{align}
&S = 4\pi {e^{ - 2{\phi _0}}}\left( {m + \sqrt {{m^2} - {q^2}} } \right)\nonumber \\
&\beta  = 2\pi \ell \frac{{\left( {m + \sqrt {{m^2} - {q^2}} } \right)}}{{\sqrt {{m^2} - {q^2}} }}\nonumber \\
&T = \frac{1}{{2\pi \ell }}\frac{{\sqrt {{m^2} - {q^2}} }}{{\left( {m + \sqrt {{m^2} - {q^2}} } \right)}}
\end{align}
The Mass and charge operators are given by:
\begin{align}
&M = \frac{4}{\ell }{e^{ - 2\phi }}\left( {\frac{{a{{\dot \phi }^2}}}{N} - \frac{{N{{\phi '}^2}}}{a}} \right) + \frac{1}{\ell }{e^{ - 2\phi }} + \frac{1}{{2\ell }}{q^2}{e^{2\phi }}\nonumber \\
&Q = {e^{ - 2\phi }}{A_t}^\prime 
\end{align}
The Hamiltonian constraint is:
\begin{equation}H =  - \frac{{{e^{2\phi }}{p_a}{p_\phi }}}{{4a}} - \frac{{{e^{2\phi }}p_a^2}}{4} + \frac{{{e^{ - 2\phi }}}}{{{\ell ^2}}} + \frac{{4{e^{ - 2\phi }}}}{{{a^2}}}\left( { - {{\phi '}^2} - \frac{{a'}}{a}\phi ' + \phi ''} \right)\end{equation}
The black hole states satisfy:
\begin{align}
&H\psi  = 0\nonumber \\
&M\psi  = m\psi \nonumber \\
&Q\psi  = q\psi 
\end{align}
To apply quantum computing one solves these eigenvalue problems using quantum algorithms which we discuss in the next section.

\section{Quantum Computing}

Quantum computing is a potentially disruptive form of computing which utilizes quantum principles of superposition and entanglement in its operations. It can potentially speedup several important algorithms and applications. One algorithm that runs on near term quantum computers is the Variational Quantum Eigensolver (VQE) \cite{Tilly:2021jem}\cite{vqe}\cite{vqe-excited}. The VQE hybrid is a quantum-classical algorithm  which has been used extensively in Quantum Chemistry to determine the lowest energy state. However the VQE can be used for other applications including Black Hole Radiation and Quantum Cosmology. In this section we use the VQE to study the interior black hole states of rotating and charged black holes. The interior black hole has features of a collapsing cosmology and thus we are able to apply quantum computing approaches used for quantum cosmology to study it. The main difference is that there is an additional operator the Mass operator that needs to be considered in additional to the Hamiltonian constraint that is used for quantum cosmology.

The VQE hybrid classical-quantum algorithm represents the Hamitonian is terms of $q$ qubits of as a $2^q \times 2^q$ matrix in terms discrete basis. We can use with the position basis where the position matrix is diagonal or the harmonic oscillator basis where the  momentum and position matrices are sparse with nonzero values one off the diagonal. These correspond to the plane wave or Gaussian basis used in chemistry. One one has represented the Hamiltonian as a finite matrix one the expand the matrix in terms of a sum of Pauli Strings 

The Hamiltonian or Mass operator is then a function of these variables and is expanded in terms of Pauli terms as:
\begin{equation}H = \sum\limits_n {{a_n}{P_n}} \end{equation}
where 
\begin{equation}{P_n} = {\sigma _i} \otimes {\sigma _j} \otimes  \ldots {\sigma _k}\end{equation}
and $i,j,k,\ldots$ go from 0 to 3.

 The VQE algorithm calculates the lowest eigenvalue and ground state of an operator by using the variational method of quantum mechanics where the variational wave functions are represented as quantum gates and the variational parameters are rotation angles for these gates. These parameters are optimized by  varying the angles in the gates and evaluating the expectation value of the operator until the process converges. This yields a least upper bound on the lowest eigenvalue given by:
\begin{equation}{m_0} \le \frac{{\left\langle {\psi ({\theta _i}} \right|M\left| {\psi ({\theta _i}} \right\rangle }}{{\left\langle {\psi ({\theta _i}} \right|\left. {\psi ({\theta _i}} \right\rangle }}\end{equation}

\subsection*{Gaussian or Simple Harmonic Oscillator basis}

The first step in quantum computing simulation is to choose a representation or basis for the Hamiltonian. 
The Gaussian or Harmonic Oscillator representation is a very useful basis based on the matrix treatment of the simple harmonic oscillator which is sparse in representing both the position and momentum operators. For the position operator we have:
\begin{equation} 
 Q_{osc} = \frac{1}{\sqrt{2}}\begin{bmatrix}
 
   0 & {\sqrt 1 } & 0 &  \cdots  & 0  \\ 
   {\sqrt 1 } & 0 & {\sqrt 2 } &  \cdots  & 0  \\ 
   0 & {\sqrt 2 } &  \ddots  &  \ddots  & 0  \\ 
   0 & 0 &  \ddots  & 0 & {\sqrt {N-1} }  \\ 
   0 & 0 &  \cdots  & {\sqrt {N-1} } & 0  \\ 
\end{bmatrix}
  \end{equation}
while for the momentum operator we have:
\begin{equation}
 P_{osc} = \frac{i}{\sqrt{2}}\begin{bmatrix}
 
   0 & -{\sqrt 1 } & 0 &  \cdots  & 0  \\ 
   {\sqrt 1 } & 0 & -{\sqrt 2 } &  \cdots  & 0  \\ 
   0 & {\sqrt 2 } &  \ddots  &  \ddots  & 0  \\ 
   0 & 0 &  \ddots  & 0 & -{\sqrt {N-1} }  \\ 
   0 & 0 &  \cdots  & {\sqrt {N-1} } & 0  \\ 
\end{bmatrix}
  \end{equation}
The Hamiltonian is then represented as $H(q,p) = H( Q_{osc}, P_{osc})$ and is expanded interms of qubits.

\subsection*{Position basis}

In the position basis the position matrix is diagonal but the momentum matrix is dense and constructed from the position operator using a Sylvester matrix $F$. In the position basis the position matrix is:
\begin{equation}
{\left( {{Q_{pos}}} \right)_{j,k}} = \sqrt {\frac{{2\pi }}{{4N}}} (2j - (N + 1)){\delta _{j,k}}
\end{equation}
and the momentum matrix is:
\begin{equation}{P_{pos}} = {F^\dag }{Q_{pos}}F\end{equation}
where 
\begin{equation}{F_{j,k}} = \frac{1}{{\sqrt N }}{e^{\frac{{2\pi i}}{{4N}}(2j - (N + 1))(2k - (N + 1))}}\end{equation}
The Hamiltonian is then represented as $H(q,p) = H( Q_{pos}, P_{pos})$

\subsection*{Multiple variables}
In this paper we will use the oscillator basis and tensor products for two variables which we denote as $(a,b)$, $(x,y)$ $(u,v)$, or $(z,w)$ depending on the type of black hole. The multiple variables can be constructed through tensor products as 
\[u = {Q_{osc}} \otimes {I_{{2^{\frac{q}{2}}}}}\]
\[v = {I_{_{{2^{\frac{q}{2}}}}}} \otimes {Q_{osc}}\]
\[{p_u} = {P_{osc}} \otimes {I_{{2^{\frac{q}{2}}}}}\]
\begin{equation}{p_v} = {I_{_{{2^{\frac{q}{2}}}}}} \otimes {P_{osc}}\end{equation}
where $q$ is the number of qubits so the matrices will be of size $2^q \times 2^q$. The Hamiltonian or Mass operator is then a function of these variables and is expanded in terms of Pauli terms as:
\begin{equation}H(u,v,{p_u},{p_v}) = \sum\limits_n {{a_n}{P_n}} \end{equation}
where 
\begin{equation}{P_n} = {\sigma _i} \otimes {\sigma _j} \otimes  \ldots {\sigma _k}\end{equation}
and $i,j,k,\ldots$ go from 0 to 3.

\subsection*{3D rotating BTZ black hole}

It is interesting that the $r \leftrightarrow t$  transformation turns the static black hole solution outside the horizon to a time dependent solution inside the horizon. This may be connected to the phenomenon of Hawking radiation and particle production from black hole decay. This contrasts with the extreme charged black hole which has zero temperature and  also does not develop a time dependent metric inside the horizon as the $g_{tt}$ and $g_rr$ components of the metric do not change sign in that case. In any event the representation if the metric inside the horizon in terms of a time dependent metric allows on to approach the quantization of the black hole interior using the same techniques as in quantum cosmology. We simplify the $H$ and $M$ operators inside the horizon using the $r \leftrightarrow t$  transformation and the mini-superspace approximation. The midi-superspace approximation involved terms in the operators that dependend on the spatial gradient and will be investigated in future research.

\begin{figure}
\centering
  \includegraphics[width = .5 \linewidth]{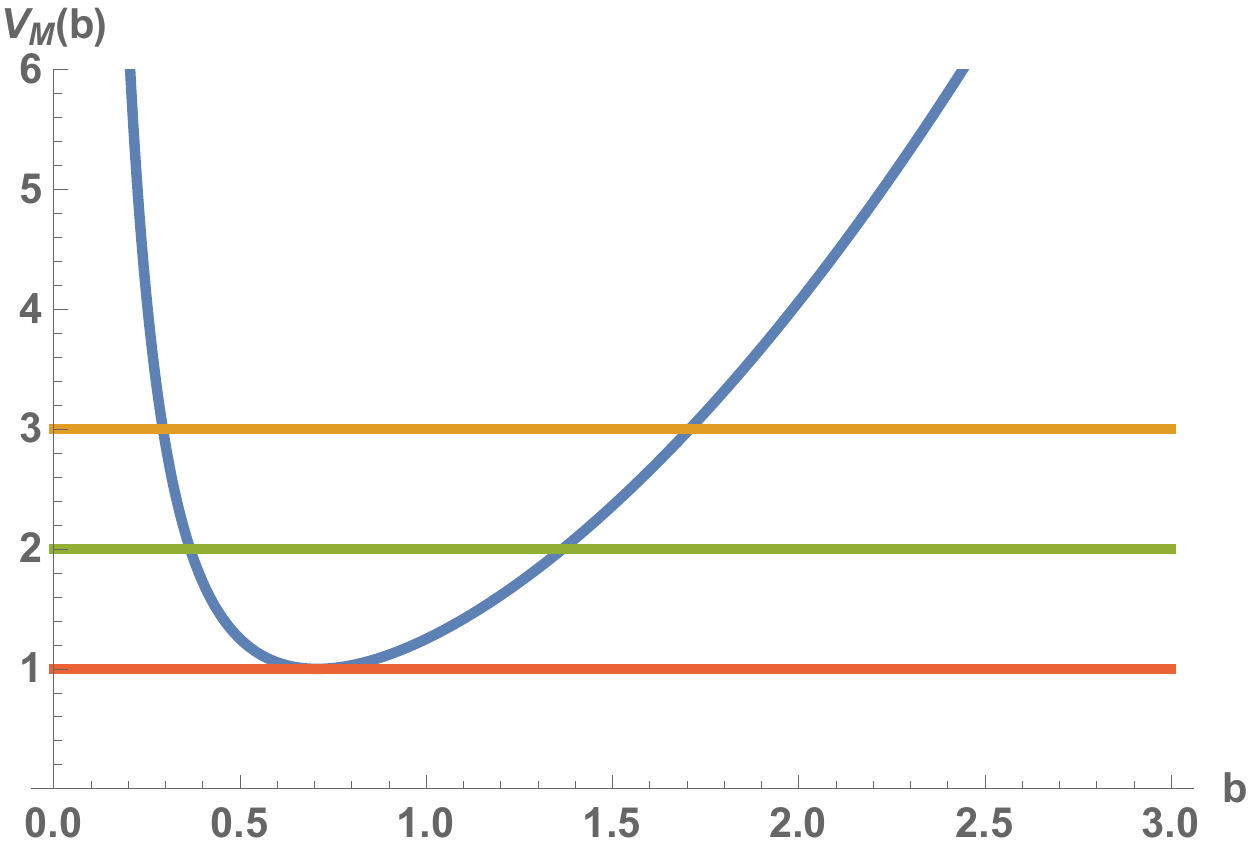}
  \caption{Plot of the mass potential for the 3D rotating BTZ black hole give by: $ V_M(b) = \frac{{{b^2}}}{{{\ell ^2}}} + \frac{{{J^2}}}{{4{b^2}}}$ for angular momentum $J=1$. There is not Nariai or upper limit to the mass in this case. The red line indicates $M=J=1$ which is an extreme rotating black hole and the lower limit of the black hole mass at this angular momentum. 
  }
  \label{fig:Radion Potential}
\end{figure}

\begin{figure}
\centering
  \includegraphics[width = .5 \linewidth]{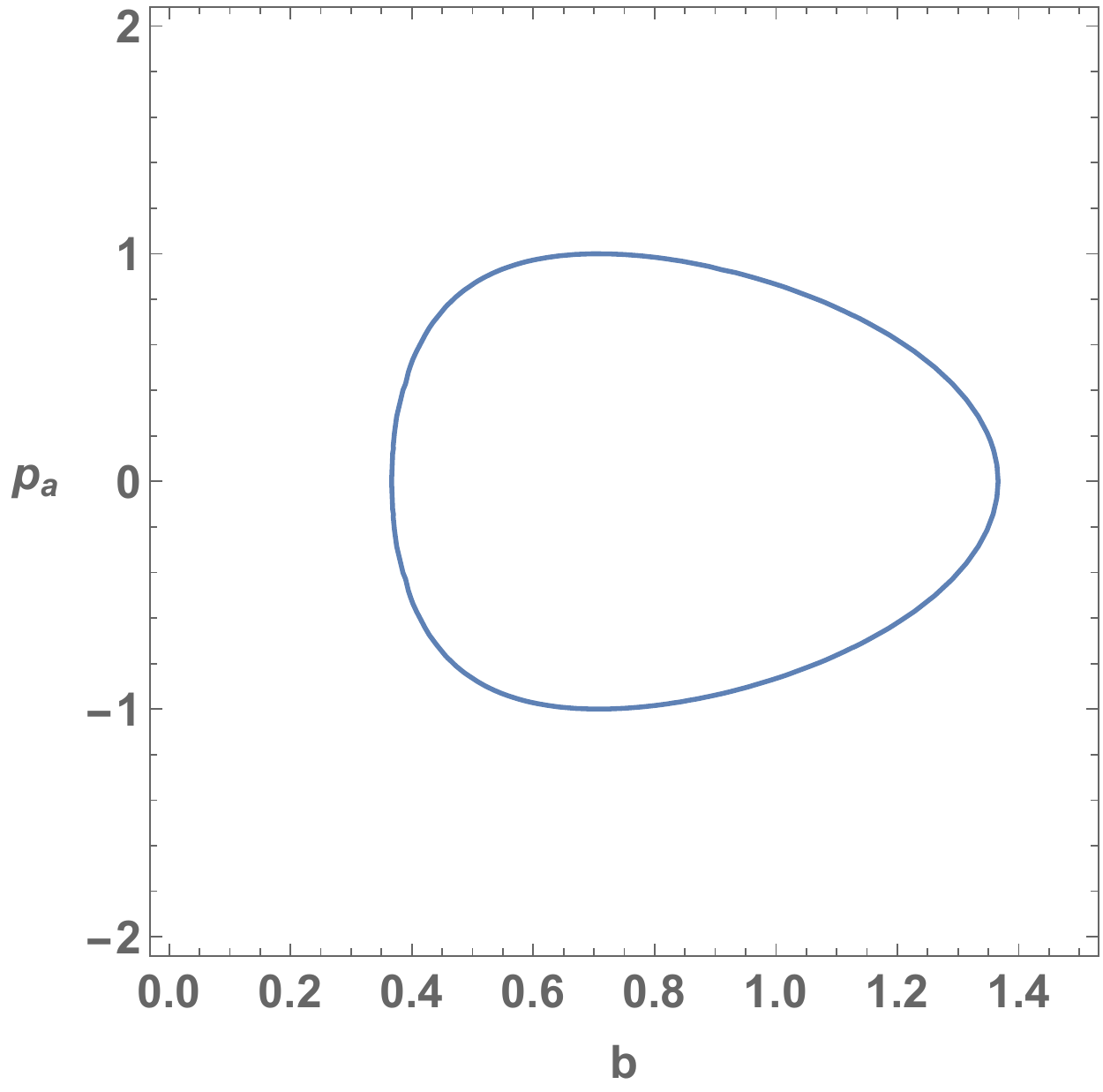}
  \caption{Contour plot for the 3D rotating BTZ black hole of $M  = p_a^2 + \frac{{{b^2}}}{{{\ell ^2}}} + \frac{{{J^2}}}{{4{b^2}}}$ for $M=2$ and $J=1$. In this case there is no region that would indicate the presence of a Nariai upper limit on the mass. 
  }
  \label{fig:Radion Potential}
\end{figure}

Concentrating on the mini-superspace expressions for the black hole interior where terms with spatial gradients are set to zero the Hamiltonian and Mass operators are given by:
\begin{align}
&H = {p_a}{p_b} + \frac{{ab}}{{{\ell ^2}}} - \frac{{{J^2}a}}{{4{b^3}}}\nonumber \\
&M = p_a^2 + \frac{{{b^2}}}{{{\ell ^2}}} + \frac{{{J^2}}}{{4{b^2}}}
\end{align}
Defining $x,y$ through:
\[a = \frac{1}{{\sqrt 2 }}(x + y)\]
\begin{equation}b = \frac{1}{{\sqrt 2 }}(x - y)\end{equation}
we have :
\[H = \frac{1}{2}\left( {p_x^2 - p_y^2} \right) + \frac{1}{2}\left( {{x^2} - {y^2}} \right) - \frac{{{J^2}\left( {x + y} \right)}}{{2{{\left( {x - y} \right)}^3}}}\]
\begin{equation}M = \frac{1}{2}{\left( {{p_x} + {p_y}} \right)^2} + \frac{1}{2}{\left( {x - y} \right)^2} + \frac{{{J^2}}}{{2{{\left( {x - y} \right)}^2}}}\end{equation}

\subsection*{4D Charged Reissner-Nordstrom }

\begin{figure}
\centering
  \includegraphics[width = .5 \linewidth]{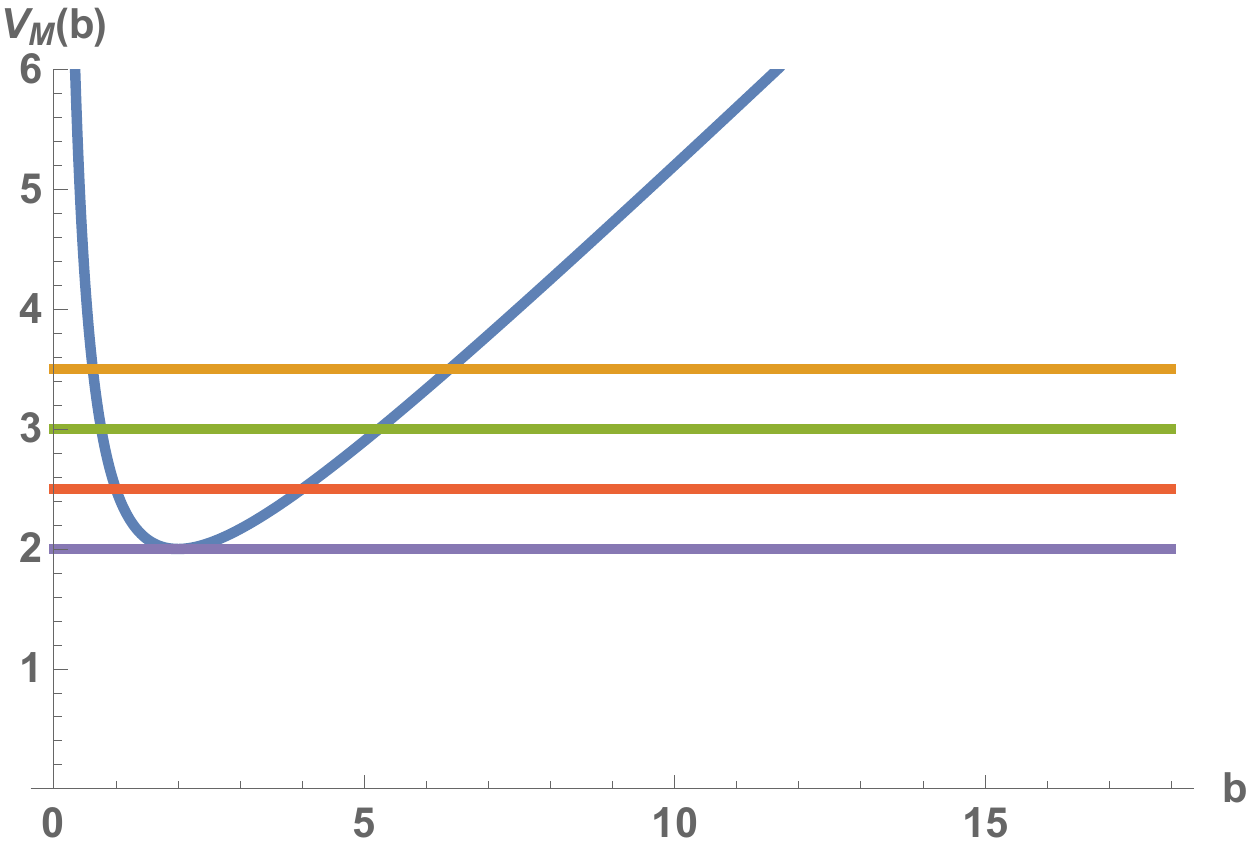}
  \caption{Plot of the mass potential for the 4D charged  RN black hole give by: $ V_M(b) = \frac{b}{2}  + \frac{{{Q^2}}}{{2b}}$ for charge momentum $Q=2$. There is not Nariai or upper limit to the mass in this case. The purple line indicates $M=Q=2$ which is an extreme charged black hole and the lower limit of the black hole mass at this charge. 
  }
  \label{fig:Radion Potential}
\end{figure}

\begin{figure}
\centering
  \includegraphics[width = .5 \linewidth]{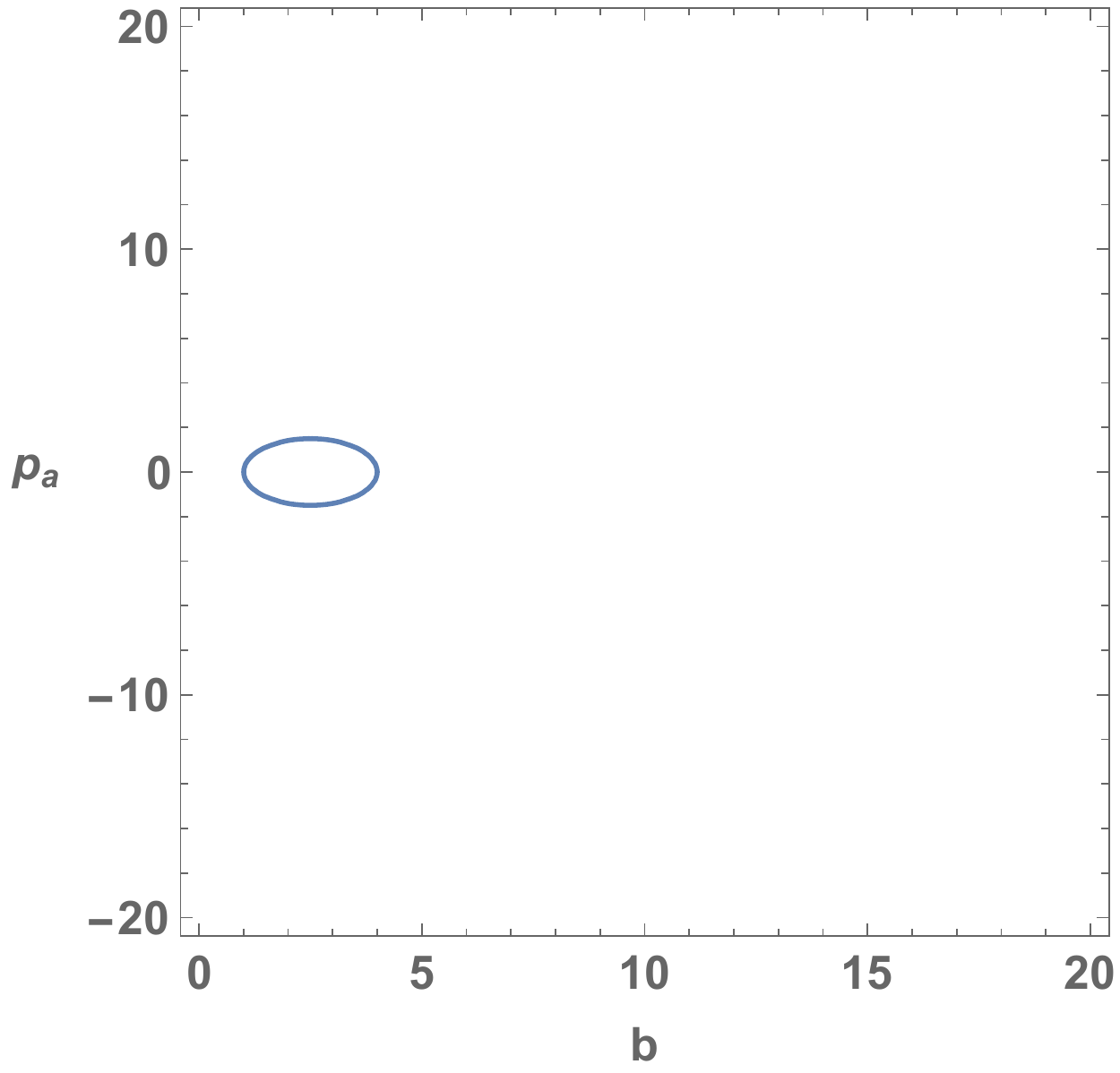}
  \caption{Contour plot for the 4D charged RN black hole of $M = \frac{p_a^2}{2b} + \frac{b}{2}  + \frac{{{Q^2}}}{{2b}}$ for $M=2.5$ and $Q=2$. In this case there is no region that would indicate the presence of a Nariai upper limit on the mass.
  }
  \label{fig:Radion Potential}
\end{figure}

For the the 4D charged Reissner-Nordstrom black hole 
the $H$ constraint and $M$ operator are:
\begin{align}
&H =   -\frac{a p_a^2}{{{2b^2}}} + \frac{p_a p_b}{b} +\frac{a}{2} - \frac{{a{Q^2}}}{{2{b^2}}}\nonumber\\
&M = \frac{p_a^2}{2b} + \frac{b}{2}  + \frac{{{Q^2}}}{{2b}}
\end{align}
For quantum computing it can be useful to use the $u,v$ representation which is defined by:
\[u = {b^{1/2}}(a + 1)\]
\begin{equation}v = {b^{1/2}}(a - 1)\end{equation}
so that
\begin{align}
&a = \frac{{u + v}}{{u - v}}\nonumber\\
&b = \frac{1}{4}{(u - v)^2}    
\end{align}
the Hamiltonian constraint $H$ and Mass operator $M$ in the $u,v$ variables are
\begin{align}
4 b H &= \frac{1}{2}\left( {p_u^2 - p_v^2} \right) + \frac{1}{2}\left( {{u^2} - {v^2}} \right)  - \frac{8Q^2 (u^2-v^2)}{(u-v)^4}\nonumber\\
4M &= \frac{1}{2}{\left( {{p_u} + {p_v}} \right)^2} + \frac{1}{2}{\left( {u - v} \right)^2} +\frac{8 Q^2}{(u-v)^2}
\end{align}
In the above the $H$ of the $a,b$ representation  was scaled by $2b$ to obtain the $H$ of the $u,v$ representation. The $M$ in the $a,b$ representation was scaled by $4$ to obtain the $M$ in the $u,v$ representation.  This form is useful when one considers the quantum computation of the Schwarzschild-de Sitter model. Physical states satisfy:
\begin{align}
&{2b H}\left| \psi  \right\rangle  = 0 \nonumber\\
&4M\left| \psi  \right\rangle = 4m_{bh} \left| \psi  \right\rangle 
\end{align}
with $m_{bh}$ the black hole mass. These are the equations we will solve using quantum computing.

\subsection*{4D Charged Reissner-Nordstrom-de Sitter}

\begin{figure}
\centering
  \includegraphics[width = .5 \linewidth]{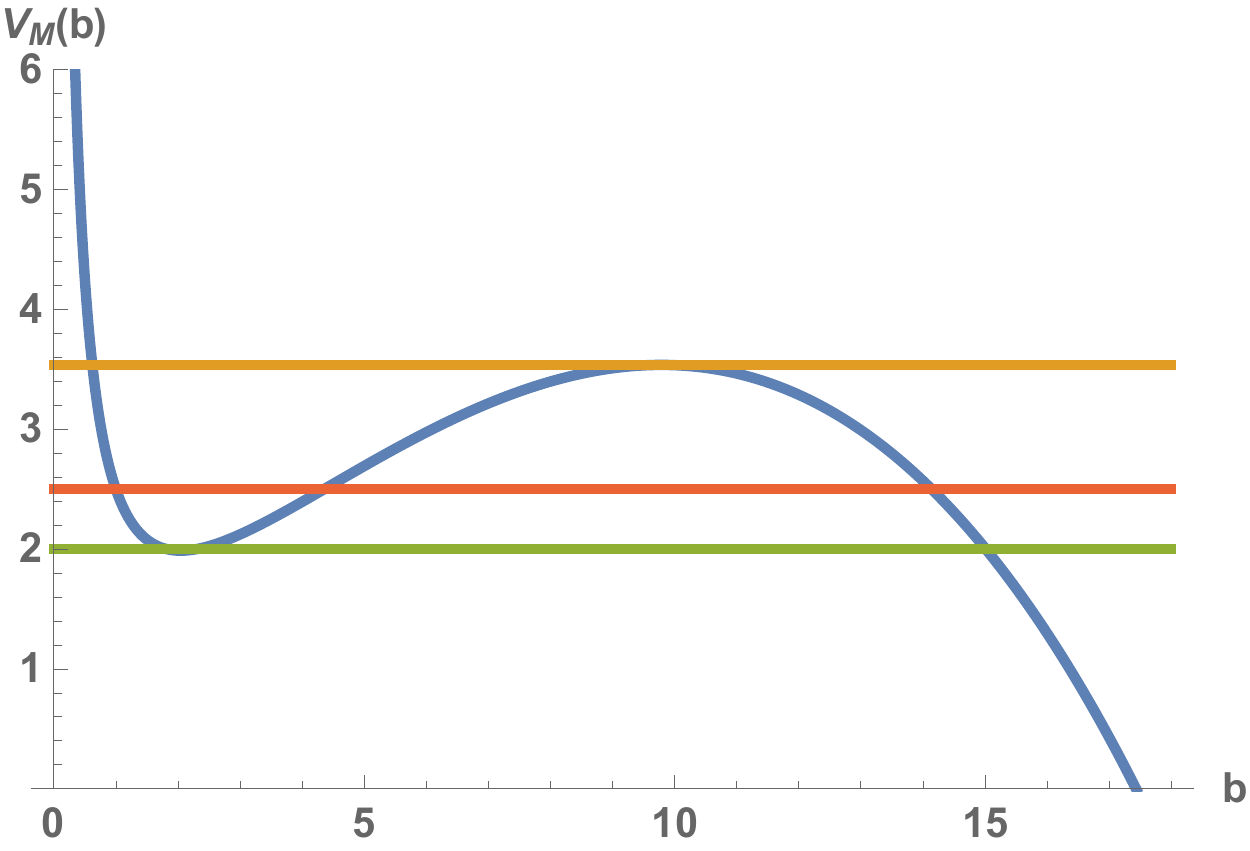}
  \caption{Plot of the mass potential for the 4D charged  RN-dS black hole give by: $ V_M(b) =  \frac{b}{2} - \frac{\lambda  b^3}{6} + \frac{{{Q^2}}}{{2b}}$ for charge $Q=2$ and $\lambda = .01$. There is a Nariai or upper limit to the mass in this case which is $M_N = 3.535433$ and denoted by the orange line. The green line indicates $M=Q=2$ which is an extreme charged black hole and the lower limit of the black hole mass at this charge.
  }
  \label{fig:Radion Potential}
\end{figure}

\begin{figure}
\centering
  \includegraphics[width = .5 \linewidth]{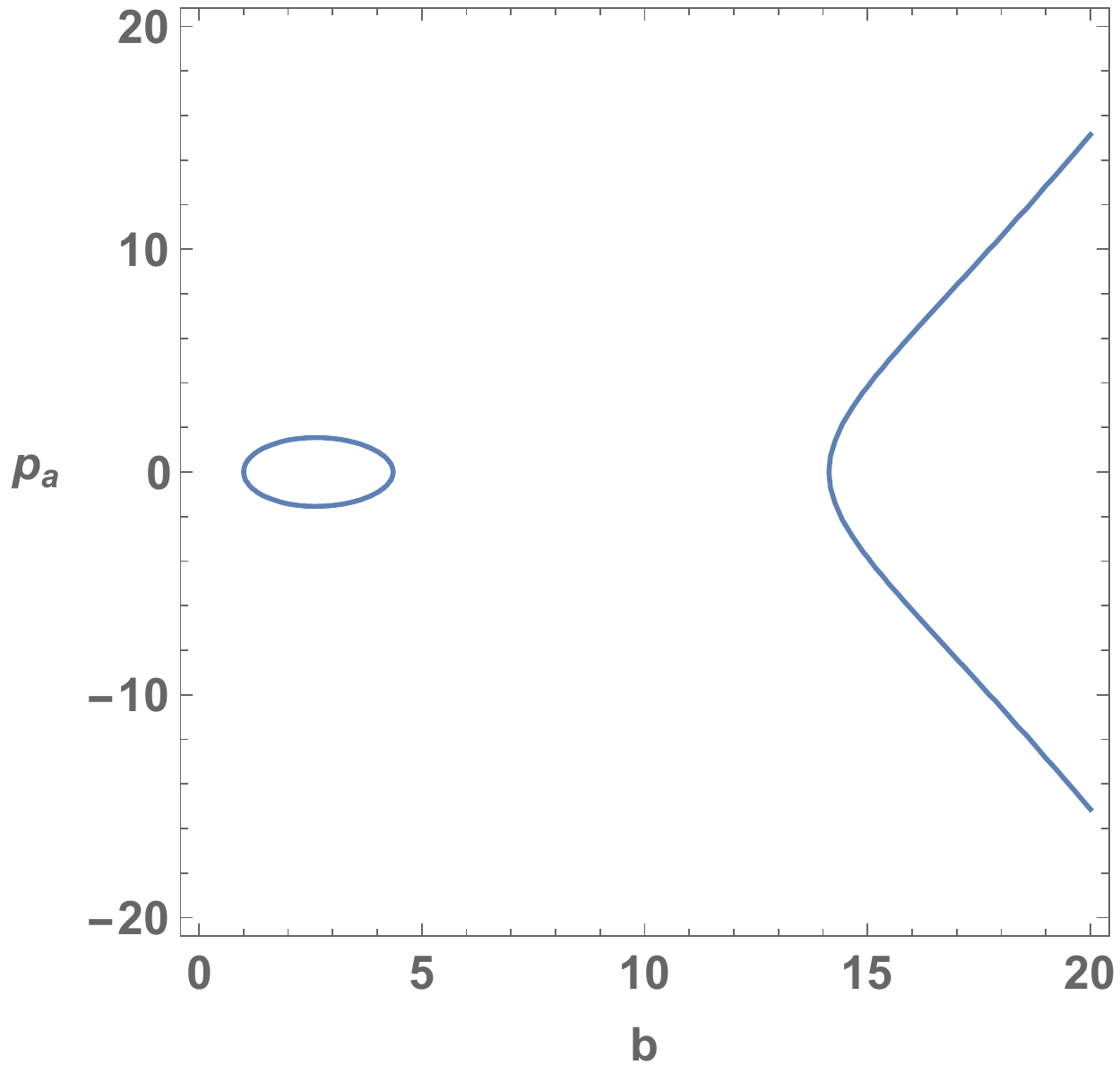}
  \caption{Contour plot for the 4D charged RN-dS black hole of $M = \frac{p_a^2}{2b} + \frac{b}{2} - \frac{\lambda  b^3}{6} + \frac{{{Q^2}}}{{2b}}$ for $M=2.5$, $\lambda = .01$ and $Q=2$. In this case there is a region to the right that  indicates the presence of a Nariai upper limit on the mass.  
  }
  \label{fig:Radion Potential}
\end{figure}

\begin{figure}
\centering
  \includegraphics[width = .5 \linewidth]{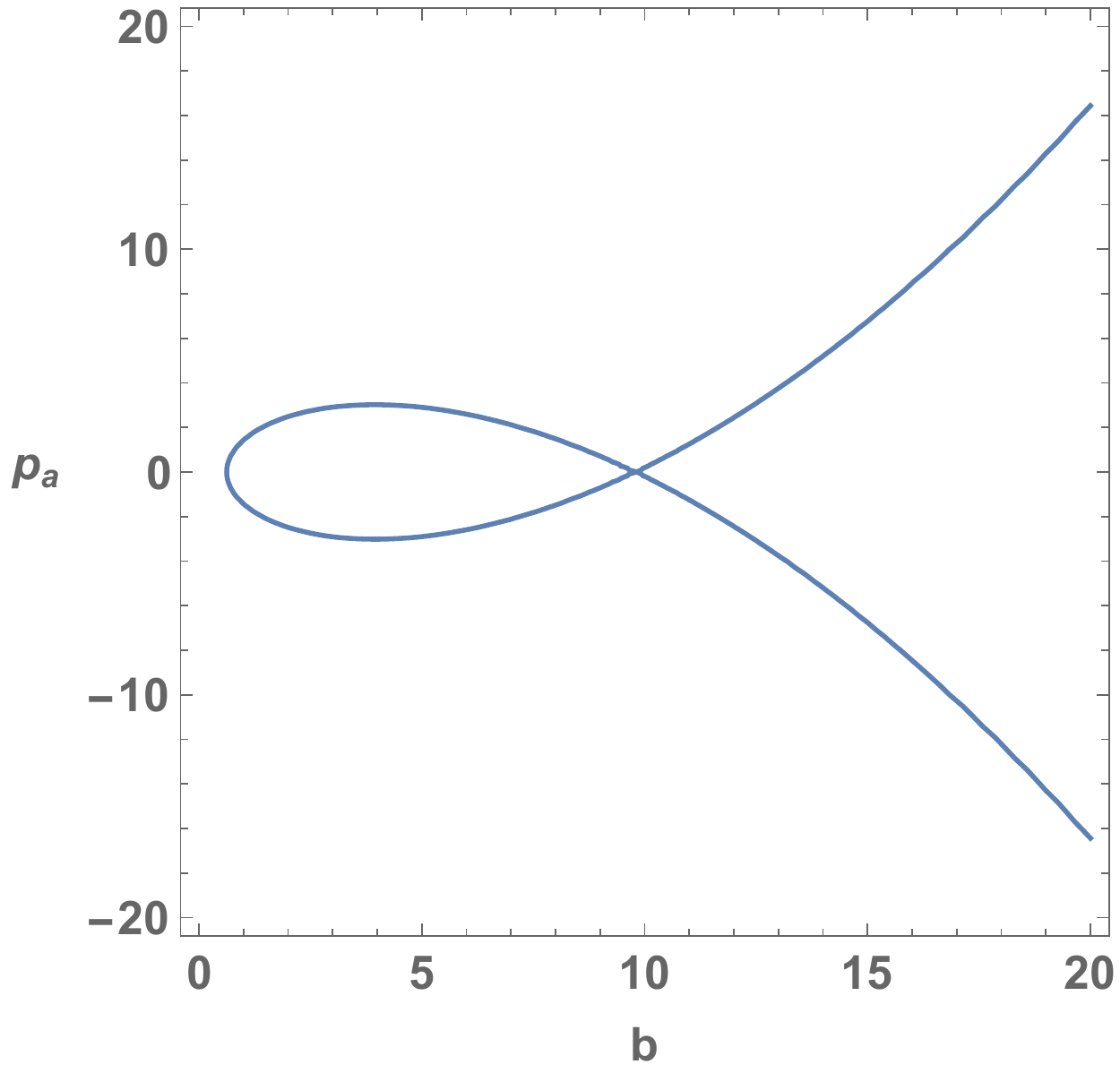}
  \caption{Contour plot for the 4D charged RN-dS black hole of $M = \frac{p_a^2}{2b} + \frac{b}{2} - \frac{\lambda  b^3}{6} + \frac{{{Q^2}}}{{2b}}$ for $M=M_N = 3.535433$, $\lambda = .01$ and $Q=2$. In this case the region to the right touches the region to the right at the Nariai point in $p_a,b$ space.
  }
  \label{fig:Radion Potential}
\end{figure}

For the 4D Charged Reissner-Nordstrom-de Sitter solution the Hamiltonian constraint and the Mass operator are:
\begin{align}
&H =   -\frac{a p_a^2}{{{2b^2}}} + \frac{p_a p_b}{b} + \frac{a}{2} - \frac{\lambda a b^2 }{2} - \frac{{a{Q^2}}}{{2{b^2}}}\nonumber\\
&M = \frac{p_a^2}{2b} + \frac{b}{2} - \frac{\lambda  b^3}{6} + \frac{{{Q^2}}}{{2b}}
\end{align}
For quantum computing it can be useful to use the $u,v$ representation which is defined by:
\[u = {b^{1/2}}(a + 1)\]
\begin{equation}v = {b^{1/2}}(a - 1)\end{equation}
so that
\begin{align}
&a = \frac{{u + v}}{{u - v}}\nonumber\\
&b = \frac{1}{4}{(u - v)^2}    
\end{align}
the Hamiltonian constraint $H$ and Mass operator $M$ in the $u,v$ variables are
\begin{align}
4 b H &= \frac{1}{2}\left( {p_u^2 - p_v^2} \right) + \frac{1}{2}\left( {{u^2} - {v^2}} \right) - \frac{\lambda }{32}\left( {{u^2} - {v^2}} \right){\left( {u - v} \right)^4} - \frac{8Q^2 (u^2-v^2)}{(u-v)^4}\nonumber\\
4M &= \frac{1}{2}{\left( {{p_u} + {p_v}} \right)^2} + \frac{1}{2}{\left( {u - v} \right)^2} - \frac{\lambda}{96}{\left( {u - v} \right)^6}+\frac{8 Q^2}{(u-v)^2}
\end{align}
In the above the $H$ of the $a,b$ representation  was scaled by $2b$ to obtain the $H$ of the $u,v$ representation. The $M$ of the $a,b$ representation was scaled by $4$ to obtain the $M$ of the $u,v$ representation.

\subsection*{2D Charged Black Hole in String Theory}

\begin{figure}
\centering
  \includegraphics[width = .5 \linewidth]{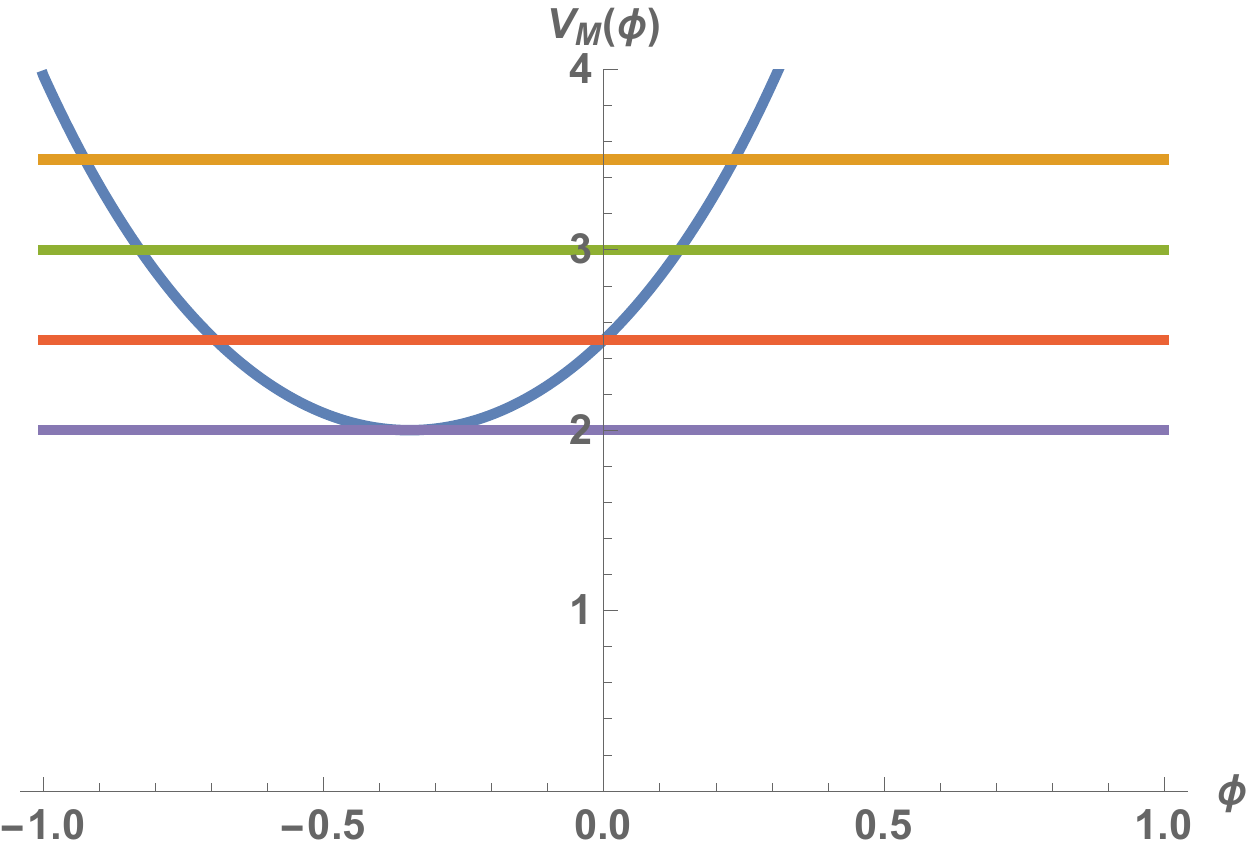}
  \caption{Plot of the mass potential for the 2D charged string black hole given by: $ V_M(b) = \frac{{{e^{ - 2\phi }}}}{{2 }} + \frac{{{Q^2}{e^{2\phi }}}}{{2 }}$ for charge $Q=2$. There is no Nariai or upper limit to the mass in this case. The purple line indicates $M=Q=2$ which is an extreme charged black hole and the lower limit of the black hole mass at this charge. 
  }
  \label{fig:Radion Potential}
\end{figure}

\begin{figure}
\centering
  \includegraphics[width = .5 \linewidth]{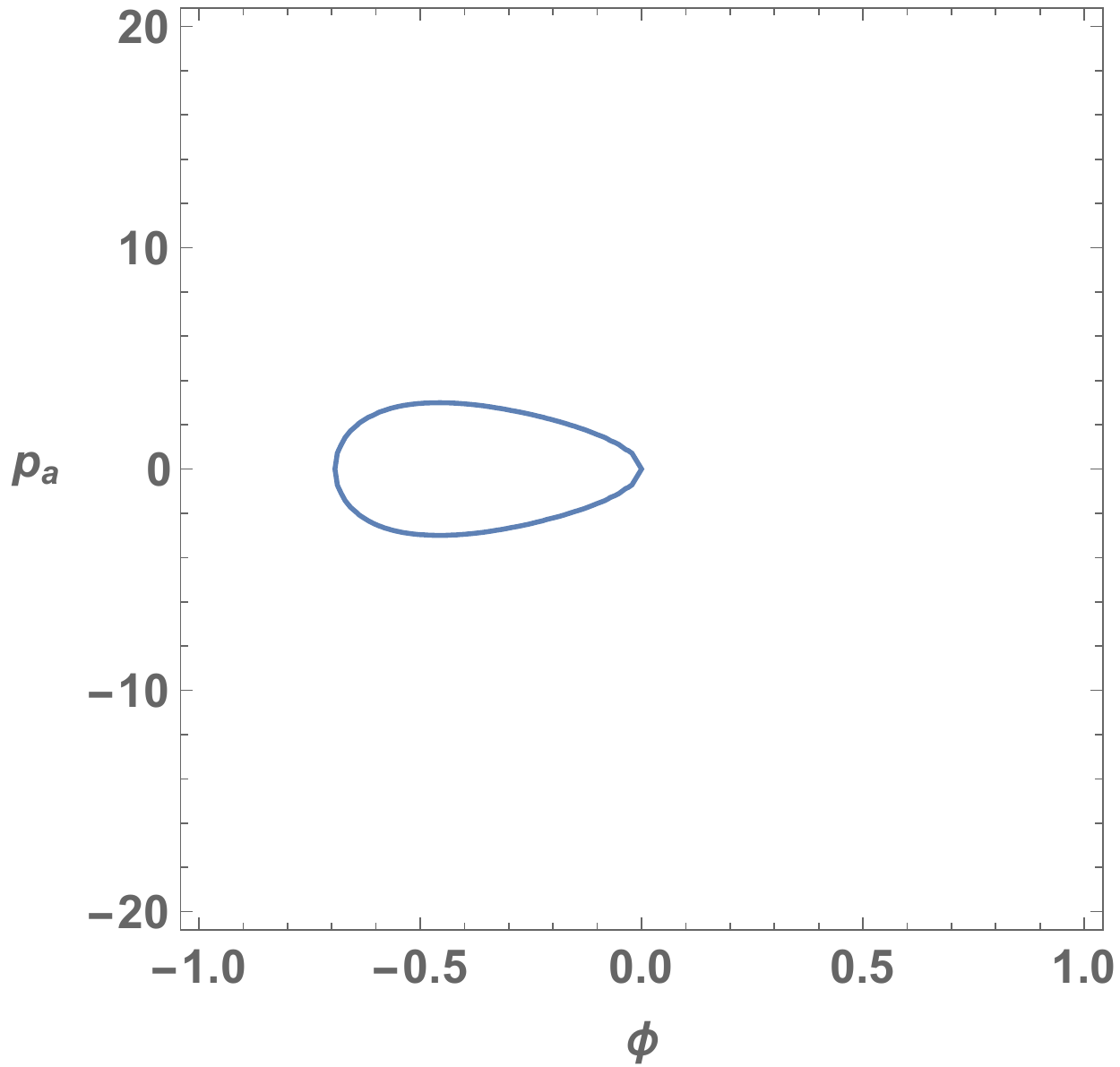}
  \caption{Contour plot for the 2D charged string black hole of $M = \frac{{{e^{2\phi }}\ell^2 p_a^2}}{8} + \frac{{{e^{ - 2\phi }}}}{{2 }} + \frac{{{q^2}{e^{2\phi }}}}{{2 }}$ for $M=2.5$ and $Q=2$. In this case there is no region that would indicate the presence of a Nariai upper limit on the mass. 
  }
  \label{fig:Radion Potential}
\end{figure}
For the 2D Charged Black Hole in String Theory
the Mass and charge operators are given by:
\begin{align}
&M = \frac{4}{\ell }{e^{ - 2\phi }}\left( {\frac{{a{{\dot \phi }^2}}}{N} - \frac{{N{{\phi '}^2}}}{a}} \right) + \frac{1}{\ell }{e^{ - 2\phi }} + \frac{1}{{2\ell }}{q^2}{e^{2\phi }}\nonumber \\
&Q = {e^{ - 2\phi }}{A_t}^\prime 
\end{align}

The Hamiltonian Constraint and Mass operators are:
\begin{align}
&H =  - \frac{{a{e^{2\phi }}p_a^2}}{4} - \frac{{{e^{2\phi }}{p_a}{p_\phi }}}{4} + \frac{{a{e^{ - 2\phi }}}}{{{\ell ^2}}}-\frac{a e^{2\phi} q^2}{2\ell^2}\nonumber \\
&M = \frac{{{e^{2\phi }}\ell^2 p_a^2}}{8} + \frac{{{e^{ - 2\phi }}}}{{2 }} + \frac{{{q^2}{e^{2\phi }}}}{{2 }}
\end{align}
defining $w,z$ variables from:
\begin{align}
&a = \frac{{w + z}}{{w - z}}\nonumber \\
&{e^{ - \phi }} = w - z
\end{align}
we have:
\begin{align}
&H = \frac{1}{{16}}\left( {p_w^2 - p_z^2} \right) + \frac{1}{{{\ell ^2}}}\left( {{w^2} - {z^2}} \right) - \frac{{{q^2}\left( {w + z} \right)}}{{{{\left( {w - z} \right)}^3}}}\nonumber \\
&M = \frac{1}{2}(\frac{{4{\ell ^2}}}{{64}}){\left( {{p_w} + {p_z}} \right)^2} + \frac{1}{2}{\left( {w - z} \right)^2} + \frac{{{q^2}}}{{2{{\left( {w - z} \right)}^2}}}
\end{align}
which is the form we will use for the quantum computations.

\section{Quantum Computing Results}

In this section, we present our results for VQE quantum computations for the four type of black holes discussed in this paper: The 3D rotating BTZ black hole, the 4D charged RN black hole, the 4D charged RN-dS black hole and the 2D charged string theory black hole.

\subsection*{3D rotating BTZ black hole}

We studied the ground state for the 3D rotating BTZ black hole. We considered states with angular momentum $J=1, 2,\dots, 5$. The ground state or state of lowest mass should be an extreme black hole of mass $M=1,2,\dots, 5$. 
\begin{equation}M = \frac{1}{2}{\left( {{p_x} + {p_y}} \right)^2} + \frac{1}{2}{\left( {x - y} \right)^2} + \frac{{{J^2}}}{{2{{\left( {x - y} \right)}^2}}}\end{equation}
We chose the Sequential Least Squares Programming (SLSQP) optimizer and the variational ansatz $R_z$ with quantum depth of three and 4 qubits so our matrices were $16\time 16$. The convergence graph for the hybrid quantum-classical computation is shown in figure . A comparison of our results to the exact value as well as the exact discrete value (obtained by replacing differential operators by finite matrices) are shown in table 2. 

\begin{figure}[htp] 
  \label{ fig7} 
  \begin{minipage}[b]{0.5\linewidth}
    \centering
    \includegraphics[width=.8\linewidth]{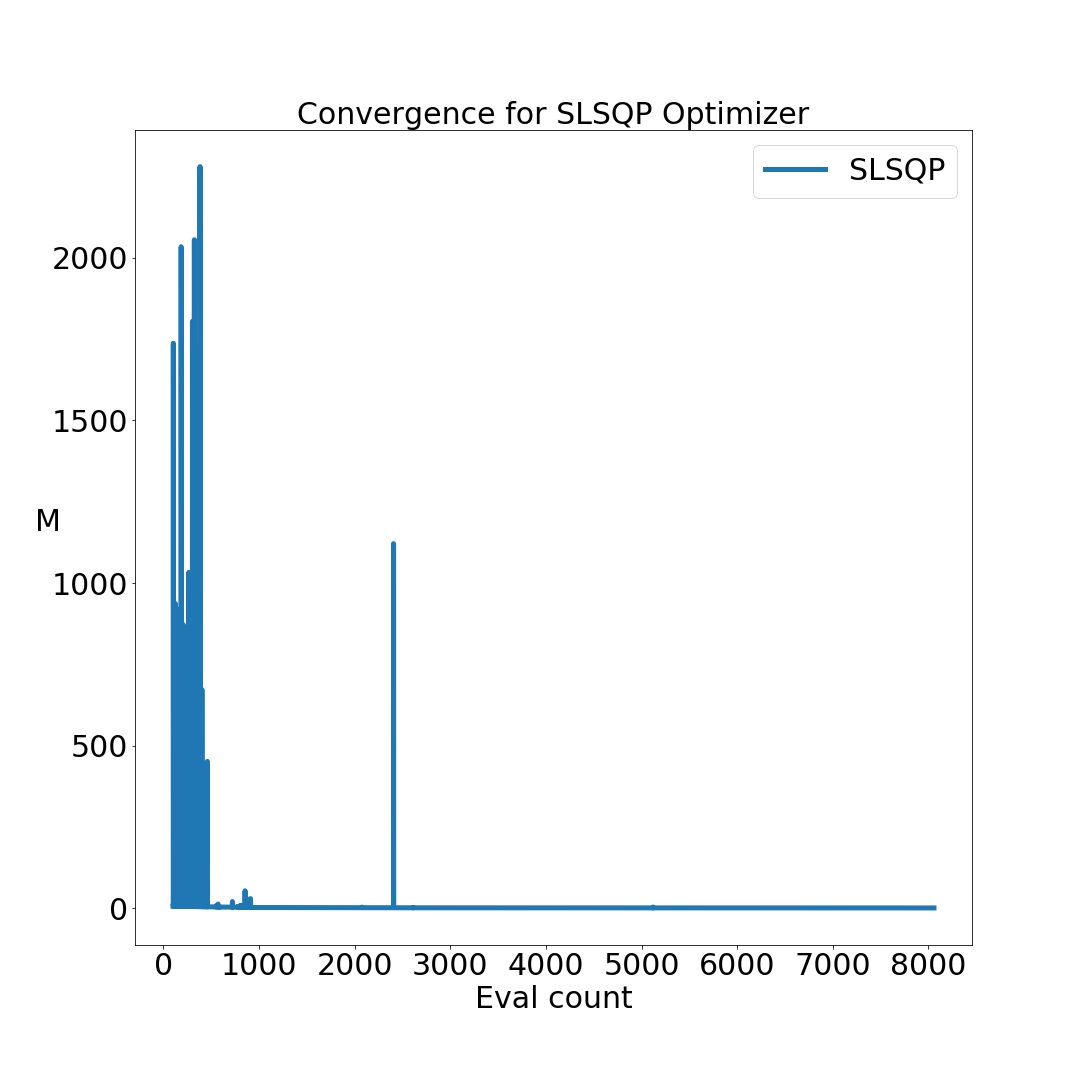}
    \vspace{4ex}
  \end{minipage}
  \begin{minipage}[b]{0.5\linewidth}
    \centering
    \includegraphics[width=.8\linewidth]{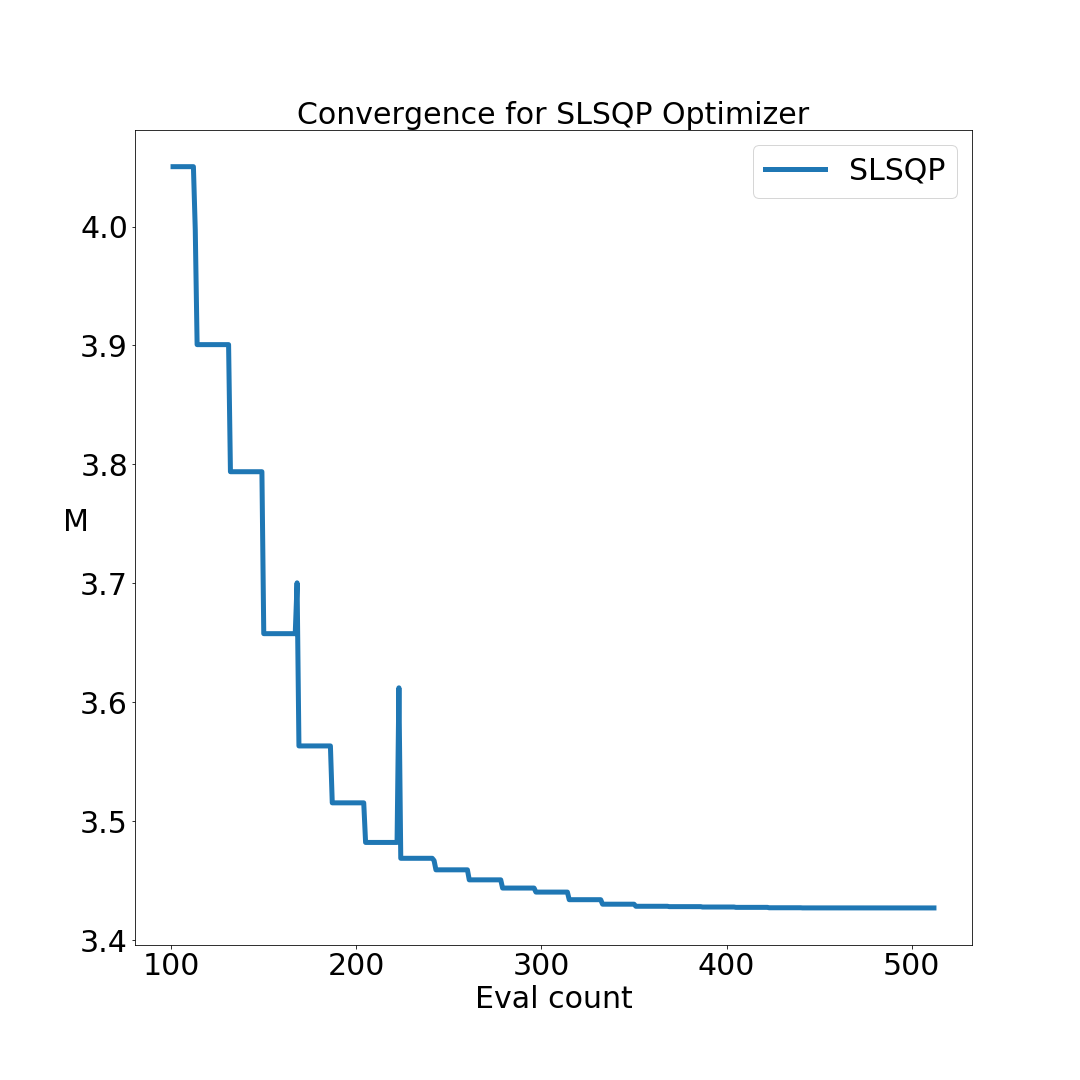} 
    \vspace{4ex}
  \end{minipage} 
  \begin{minipage}[b]{0.5\linewidth}
    \centering
    \includegraphics[width=.8\linewidth]{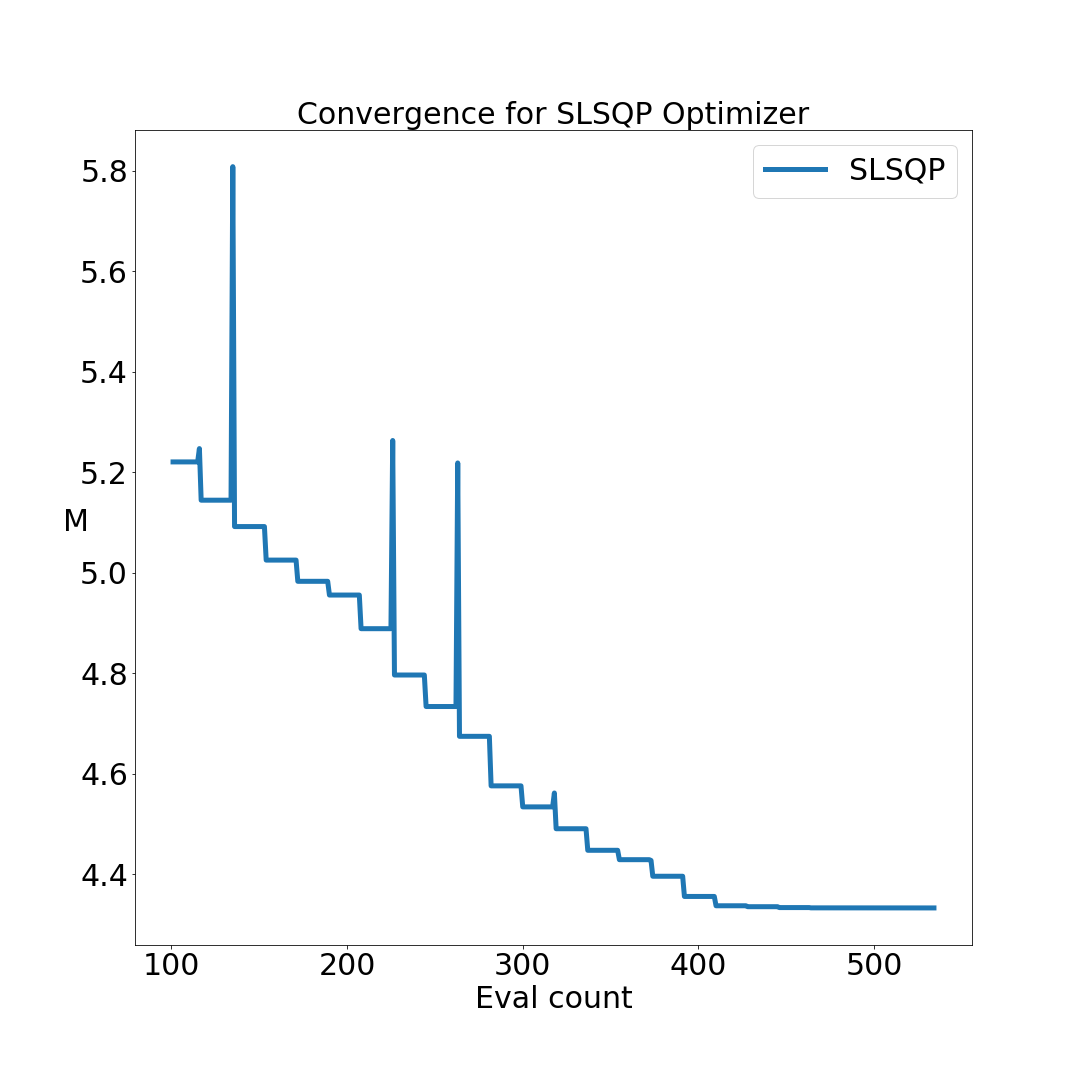} 
    \vspace{4ex}
  \end{minipage}
  \begin{minipage}[b]{0.5\linewidth}
    \centering
    \includegraphics[width=.8\linewidth]{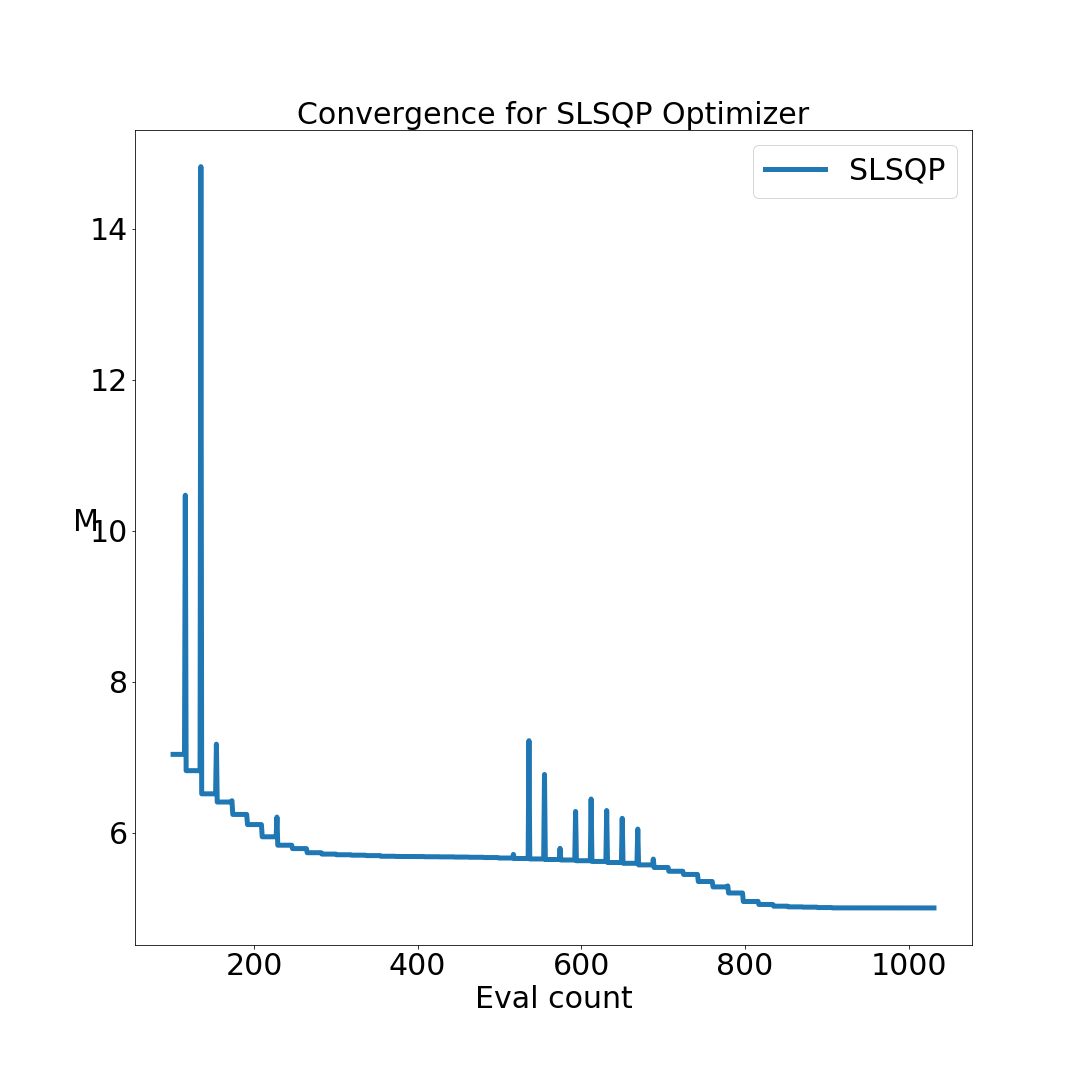} 
    \vspace{4ex}
  \end{minipage} 
    \caption{Convergence graph for the quantum VQE computation of the lowest Mass eigenstate for the 3D Rotating BTZ black hole with (upper left) $J=1$ with oscillator basis, (upper right)  $J=3$ with position basis, (lower left)  $J=4$ with position basis, (lower right)  $J=5$ with position basis. 
    }
\end{figure}



\begin{table}[ht]
\centering
\begin{tabular}{|l|l|l|l|l|l|l|l|}
\hline
Rotating BTZ  & basis &      Qubits  &  Paulis  & Exact Result & Exact Discrete & VQE Result \\ \hline
M with $J=1$ & osc & 4 & 57  & $1.0$ &  $1.09727945$ & $1.09800168$ \\ \hline
M with $J=2$ & pos & 4 & 21  & $2.0$ &  $2.00235705$ & $2.17770207$ \\\hline
M with $J=3$ & pos & 4 & 21  & $2.0$ &  $3.24533444$ & $3.42720242$ \\\hline
M with $J=4$ & pos & 4 & 21  & $4.0$ &  $4.33327732$ & $4.33327745$ \\\hline
M with $J=5$ & pos & 4 & 21  & $5.0$ &  $5.00359995$ & $5.00360008$ \\\hline

\end{tabular}
\caption{\label{tab:BasisCompare}  VQE results for the lowest eigenvalue of the Mass operator for 3D rotating BTZ  black hole using the oscillator basis. The Hamiltonian was mapped to 4, 6 and 8-qubit operators for $(x,y)$ coordinates in mini-superspace. The quantum circuit for each simulation utilized an \(R_y\) variational form, with a fully entangled circuit of depth 3. The backend used was a state-vector simulator. The VQE results were obtained using the state-vector simulator with no noise and the Sequential Least Squares Programming (SLSQP) optimizer.}
\end{table}

\subsection*{4D charged RN black hole}

We studied the ground state for the 4D charged RN black hole. We considered states with angular momentum $Q=1,2$. The ground state or state of lowest mass should be an extreme black hole of mass $M=1,2$.

Again we chose the Sequential Least Squares Programming (SLSQP) optimizer and the variational ansatz $R_z$ with quantum depth of three and 4 qubits so our matrices were $16\time 16$. The convergence graph for the hybrid quantum-classical VQE computation is shown in figure . A comparison of our results to the exact value as well as the exact discrete value (obtained by replacing differential operators by finite matrices) are shown in table 3.
\begin{equation}4M = \frac{1}{2}{\left( {{p_u} + {p_v}} \right)^2} + \frac{1}{2}{\left( {u - v} \right)^2} +\frac{8 Q^2}{(u-v)^2}\end{equation}
\begin{table}[ht]
\centering
\begin{tabular}{|l|l|l|l|l|l|l|}
\hline
Charged RN   &basis     & Qubits  &  Paulis  & Exact Result & Exact Discrete & VQE Result \\ \hline
M with $Q=1$ & osc& 4 & 57  & $1.0$ &  $1.05957665$ & $1.06250010$ \\ \hline
M with $Q=2$ & pos & 4 & 21  & $2.0$ &  $2.03869093$ & $2.03869130$ \\ \hline

\end{tabular}
\caption{\label{tab:BasisCompare}  VQE results for the lowest eigenvalue of the Mass operator for 4D charged RN  black hole using the oscillator basis. The Hamiltonian was mapped to 4, 6 and 8-qubit operators for $(u,v)$ coordinates in mini-superspace. The quantum circuit for each simulation utilized an \(R_y\) variational form, with a fully entangled circuit of depth 3. The backend used was a state-vector simulator. The VQE results were obtained using the state-vector simulator with no noise and the Sequential Least Squares Programming (SLSQP) optimizer.}
\end{table}



\begin{figure}[!htb]
\centering
\minipage{0.5\textwidth}
  \includegraphics[width=\linewidth]{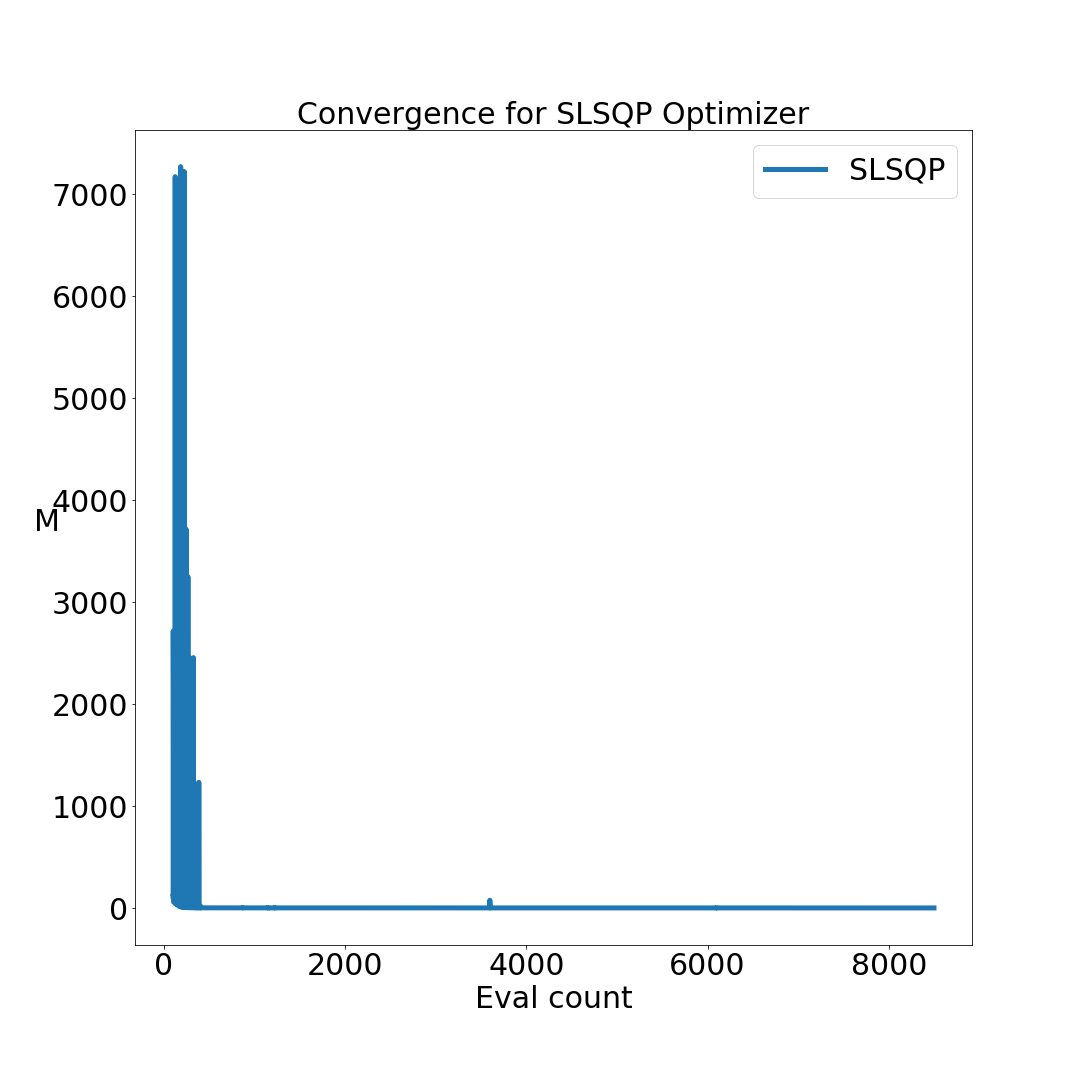}
\endminipage\hfill
\minipage{0.5\textwidth}
  \includegraphics[width=\linewidth]{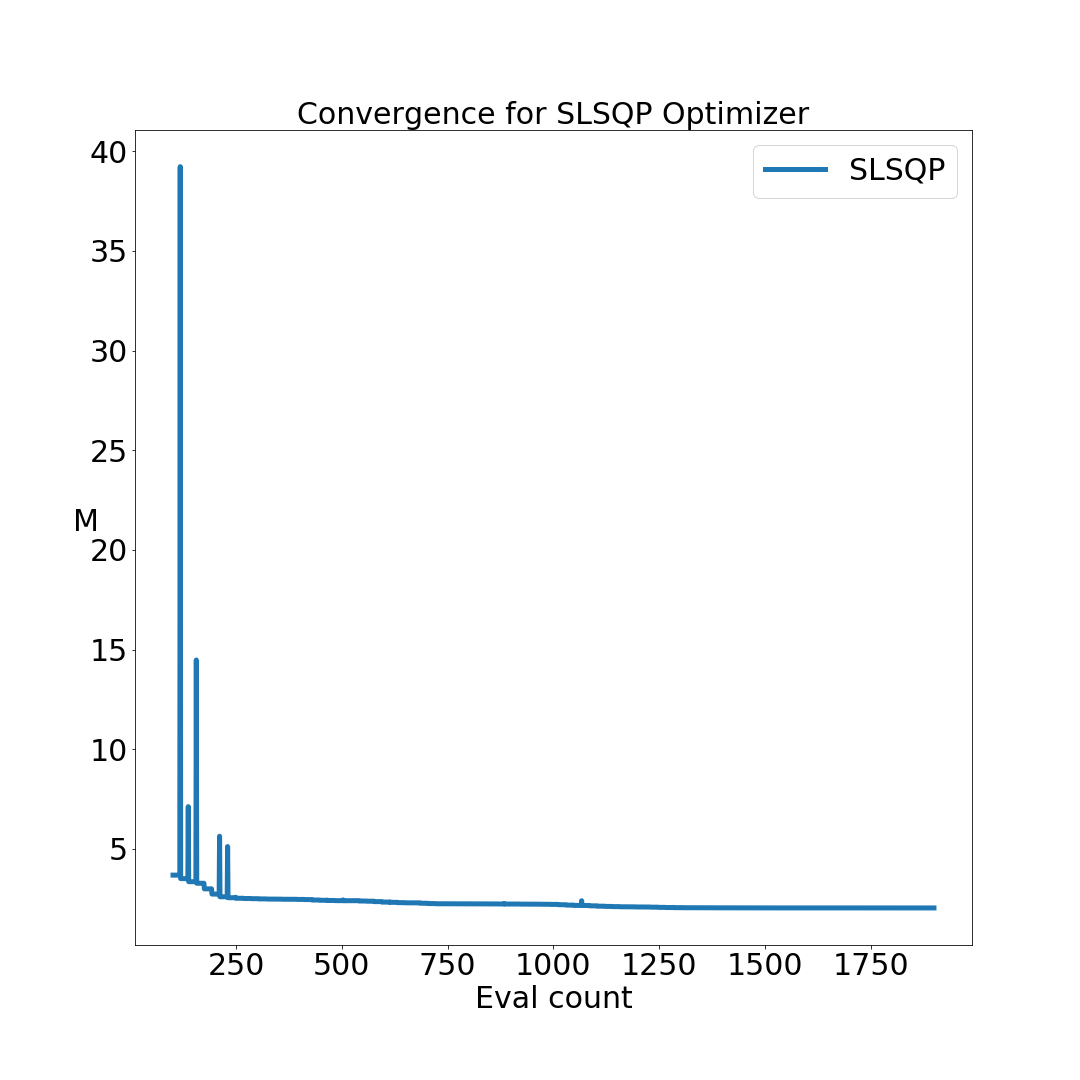}
\endminipage\hfill
\caption{Convergence graph for the quantum VQE computation of the lowest Mass eigenstate for the 4D Charged RN black hole with (left) $Q=1$ using the oscillator basis and (right) $Q=2$ using the position basis. 
}
\end{figure}



\subsection*{4D charged RN-dS black hole}

We studied the ground state for the 4D charged RN-dS black hole. We considered states with angular momentum $Q=1, 2$ with positive cosmological constant $\lambda = .01$. The ground state or state of lowest mass should be an extreme black hole of mass $M=1, 2$. 



Again we chose the Sequential Least Squares Programming (SLSQP) optimizer and the variational ansatz $R_z$ with quantum depth of three and 4 qubits so our matrices were $16\time 16$. The convergence graph for the hybrid quantum-classical computation is shown in figure . A comparison of our results to the exact value as well as the exact discrete value (obtained by replacing differential operators by finite matrices) are shown in table 4.
\begin{equation}4M = \frac{1}{2}{\left( {{p_u} + {p_v}} \right)^2} + \frac{1}{2}{\left( {u - v} \right)^2} - \frac{\lambda}{96}{\left( {u - v} \right)^6}+\frac{8 Q^2}{(u-v)^2}\end{equation}

\begin{table}[ht]
\centering
\begin{tabular}{|l|l|l|l|l|l|l|}
\hline
RNdS $\lambda = .01$  & basis    & Qubits  &  Paulis  & Exact Result & Exact Discrete & VQE Result \\ \hline
M with $Q=1$  & osc & 4 & 57  & $1.0$ &  $1.05659733$ & $1.05796099$ \\ \hline
M with $Q=2$ & pos & 4 & 21  & $2.0$ &  $ 2.03223129$ & $ 2.03223171$ \\ \hline

\end{tabular}
\caption{\label{tab:BasisCompare}  VQE results for the lowest eigenvalue of the Mass operator for 4D charged RN-dS  black hole using the oscillator and position  basis. 
}
\end{table}


\begin{figure}[!htb]
\centering
\minipage{0.5\textwidth}
  \includegraphics[width=\linewidth]{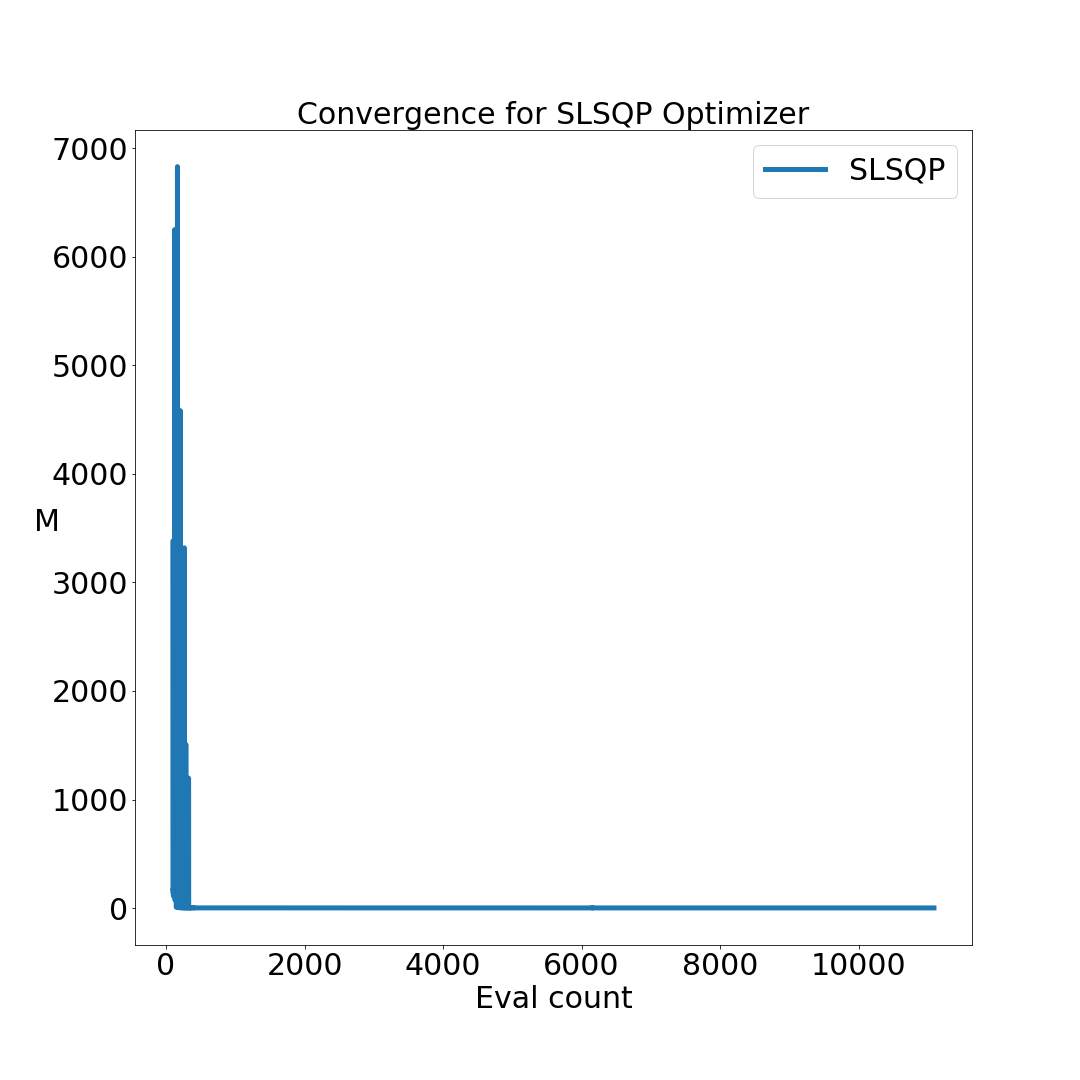}
\endminipage\hfill
\minipage{0.5\textwidth}
  \includegraphics[width=\linewidth]{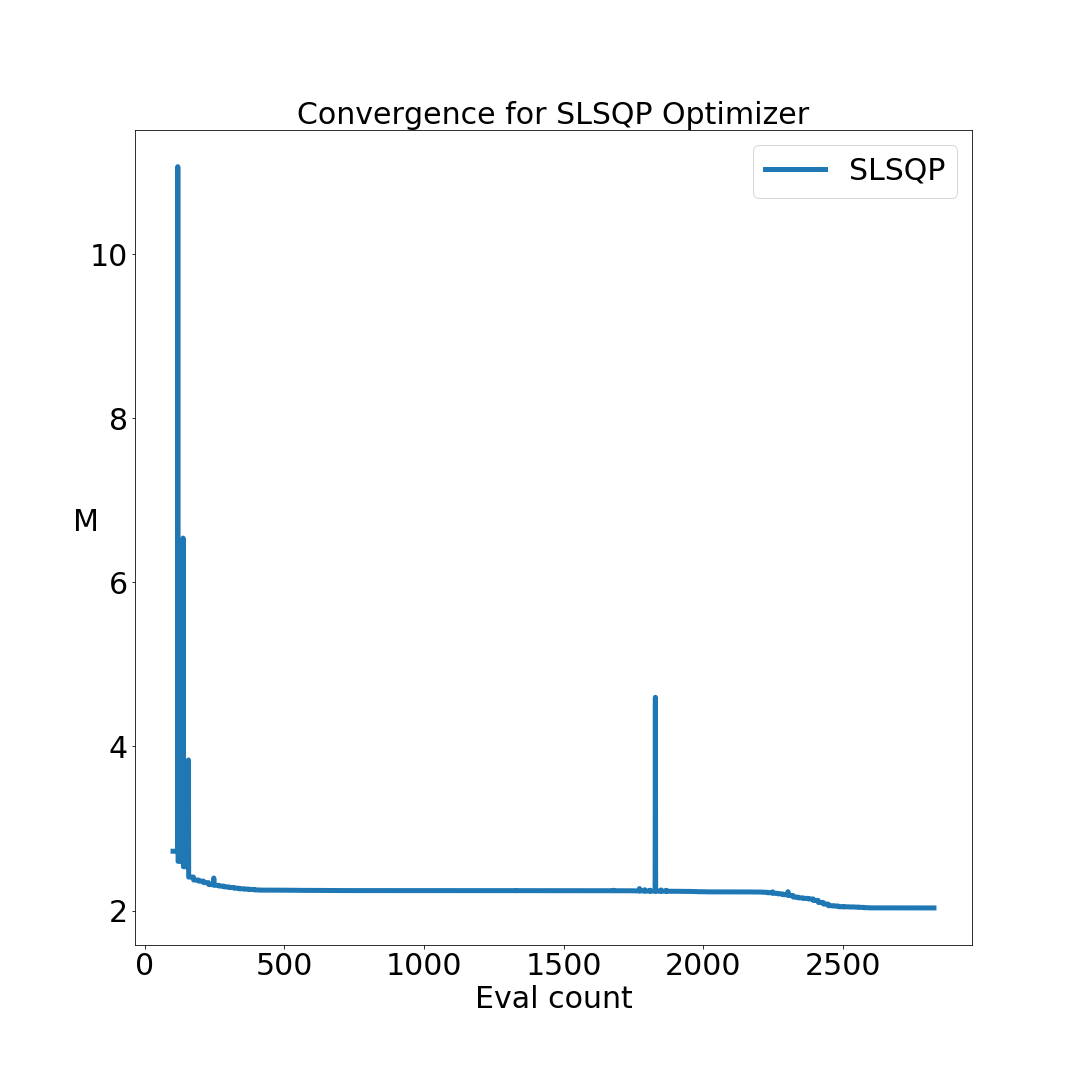}
\endminipage\hfill
\caption{Convergence graph for the quantum VQE computation of the lowest Mass eigenstate for the 4D Charged RN-dS black hole with (left) $Q=1$ using the oscillator basis and (right) $Q=2$ using the position basis. 
}
\end{figure}


\subsection*{2D charged string black hole}

We studied the ground state for the 2D charged string black hole. We considered states with angular momentum $Q=1,2$. The ground state or state of lowest mass should be an extreme black hole of mass $M=1,2$. We record our results in table 5. We see that both the oscillator and position basis yeild accurate results compared to exact discrete results. Further progress towards the exact result of the extreme charged 2D string black hole can be expected using larger number of qubits and more refined variational ansatz, however in that case this will lead to greater complexity of the quantum circuit. The $H$ and $M$ operators are:
\begin{align}
&H = \frac{1}{{16}}\left( {p_w^2 - p_z^2} \right) + \frac{1}{{{\ell ^2}}}\left( {{w^2} - {z^2}} \right) - \frac{{{q^2}\left( {w + z} \right)}}{{{{\left( {w - z} \right)}^3}}}\nonumber \\
&M = \frac{1}{2}(\frac{{4{\ell ^2}}}{{64}}){\left( {{p_w} + {p_z}} \right)^2} + \frac{1}{2}{\left( {w - z} \right)^2} + \frac{{{q^2}}}{{2{{\left( {w - z} \right)}^2}}}
\end{align}
Our results for the 2D Charged Black Hole in String Theory are given in table 5.
\begin{table}[ht]
\centering
\begin{tabular}{|l|l|l|l|l|l|l|}
\hline
Charged String & basis      & Qubits  &  Paulis  & Exact Result & Exact Discrete & VQE Result \\ \hline
M with $Q=1$ & pos & 4 & 21  & $1.0$ &  $1.16279070$ & $1.16279081$ \\ \hline
M with $Q=2$ & osc  & 4 & 57  & $2.0$ &  $2.33625205$ & $2.34228488$ \\ \hline

\end{tabular}
\caption{\label{tab:BasisCompare}  VQE results for the lowest eigenvalue of the Mass operator for 2D charged string  black hole using the oscillator basis. The Hamiltonian was mapped to 4-qubit operators for $(w,z)$ coordinates in mini-superspace. The quantum circuit for each simulation utilized an \(R_y\) variational form, with a fully entangled circuit of depth 3. The backend used was a state-vector simulator. The VQE results were obtained using the state-vector simulator with no noise and the Sequential Least Squares Programming (SLSQP) optimizer.}
\end{table}


Beside the Mass eigenvalue we were able to use the VQE method to compute the magnitude of the Hamiltonian constraint as well as the magnitude of the commutator of the Hamiltonian constraint and Mass operator with results in tables 6 and 7. In all cases we found good agreement with the expected value of zero. Similar computations of expectation values gauge constraints for gauge theories were considered in \cite{Han:2021our} using neural nets instead of the VQE method.



\begin{figure}[!htb]
\centering
\minipage{0.5\textwidth}
  \includegraphics[width=\linewidth]{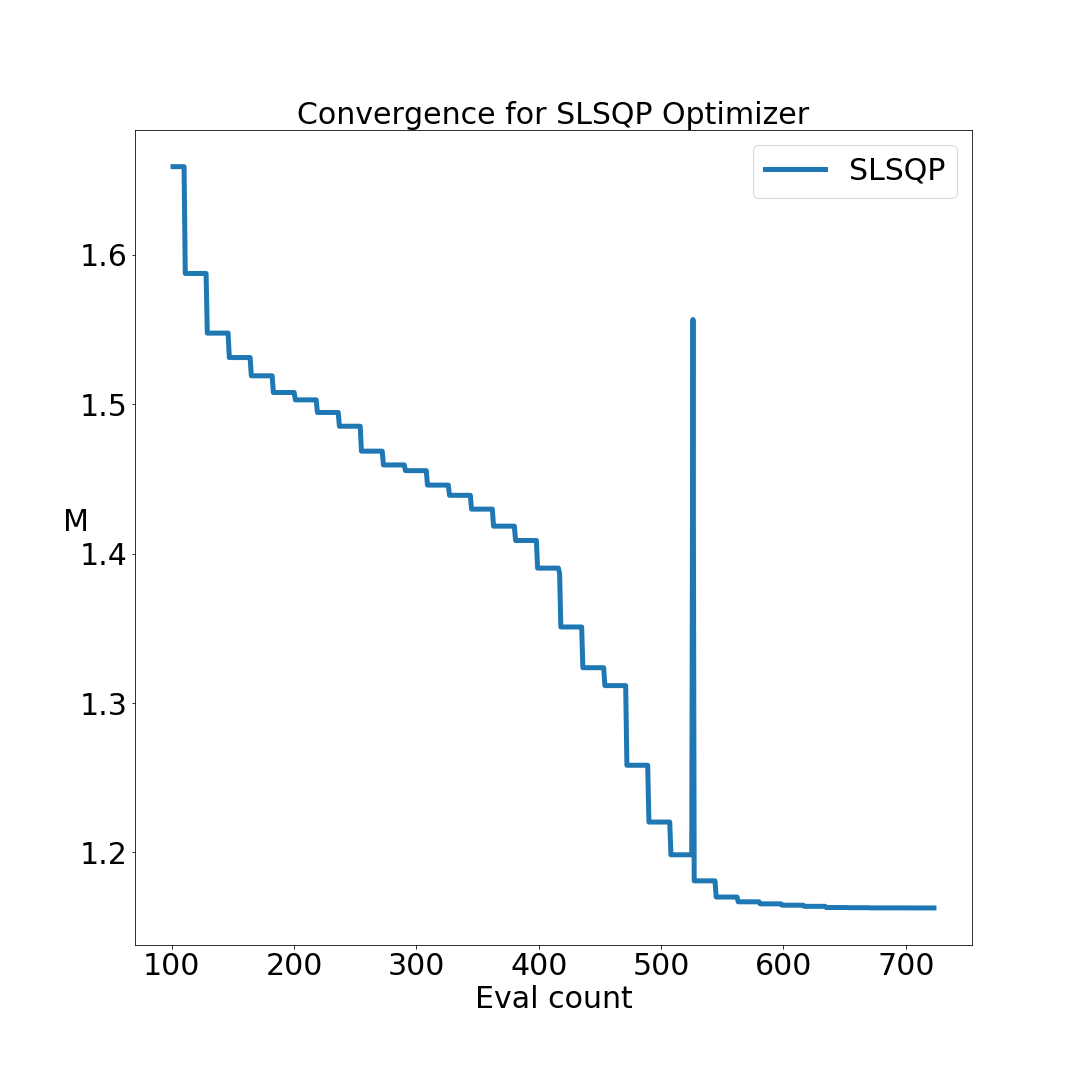}
\endminipage\hfill
\minipage{0.5\textwidth}
  \includegraphics[width=\linewidth]{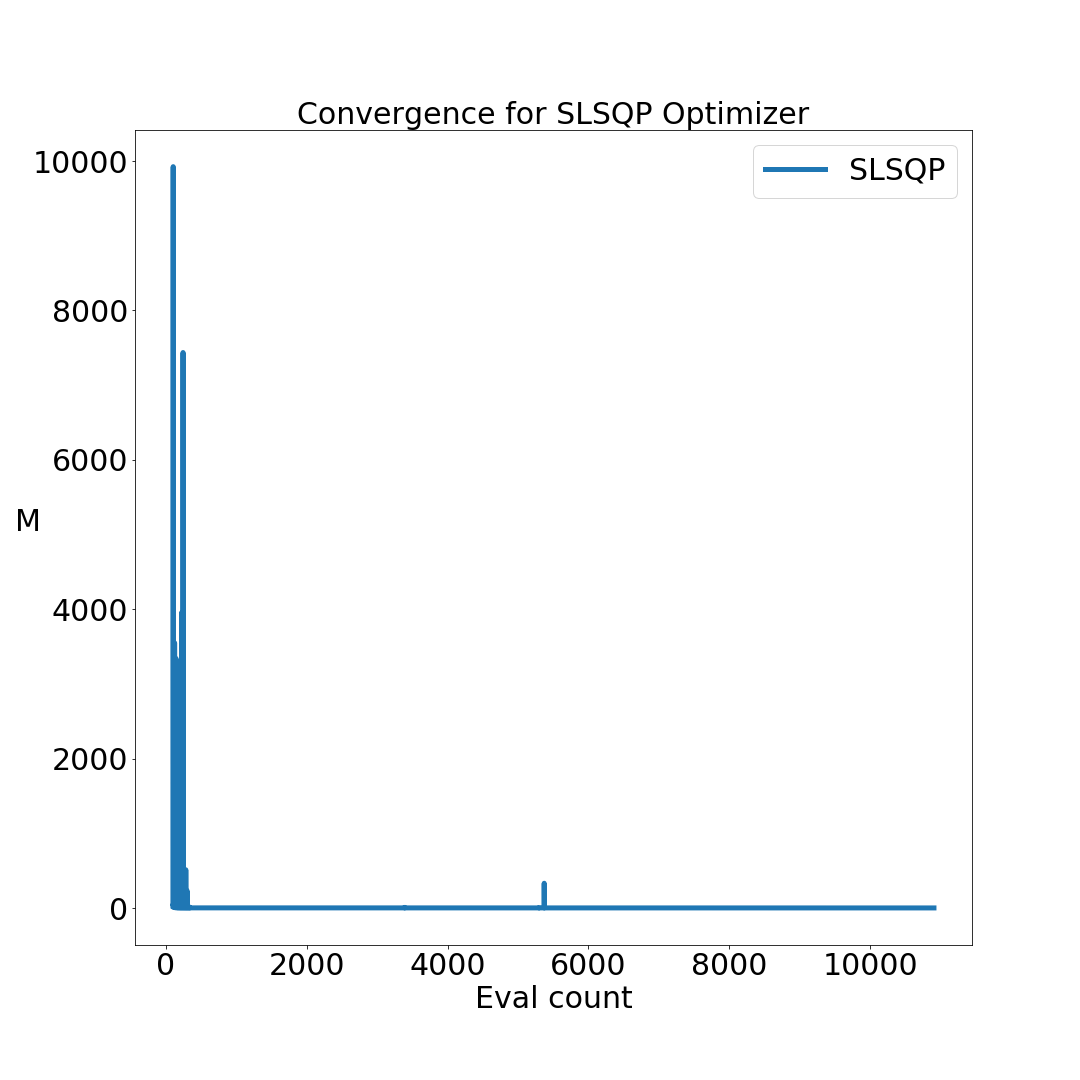}
\endminipage\hfill
\caption{Convergence graph for the quantum VQE computation of the lowest Mass eigenstate for the 2D Charged String black hole with (left) $Q=1$ using the position basis, (right)  $Q=2$ using the oscillator basis. 
}
\end{figure}



\begin{table}[ht]
\centering
\begin{tabular}{|l|l|l|l|l|l|l|}
\hline
Charged String & basis      & Qubits  &  Paulis  & Exact & Exact Discrete & VQE Result \\ \hline
$\abs{H}$ with $Q=1$ & pos & 4 & 36  & $0.0$ &  $0.0$ & $3.40074443 \times 10^{-7}$ \\ \hline
$\abs{H}$ with $Q=2$ & osc  & 4 & 122  & $0.0$ &  $0.0$ & $5.23075314\times 10^{-6}$ \\ \hline

\end{tabular}
\caption{\label{tab:BasisCompare}  VQE results for the lowest eigenvalue of the absolute value of the Hamiltonian constraint operator for 2D charged string  black hole using the oscillator basis. The Hamiltonian was mapped to 4-qubit operators for $(w,z)$ coordinates in mini-superspace. }
\end{table}

\begin{figure}[!htb]
\centering
\minipage{0.5\textwidth}
  \includegraphics[width=\linewidth]{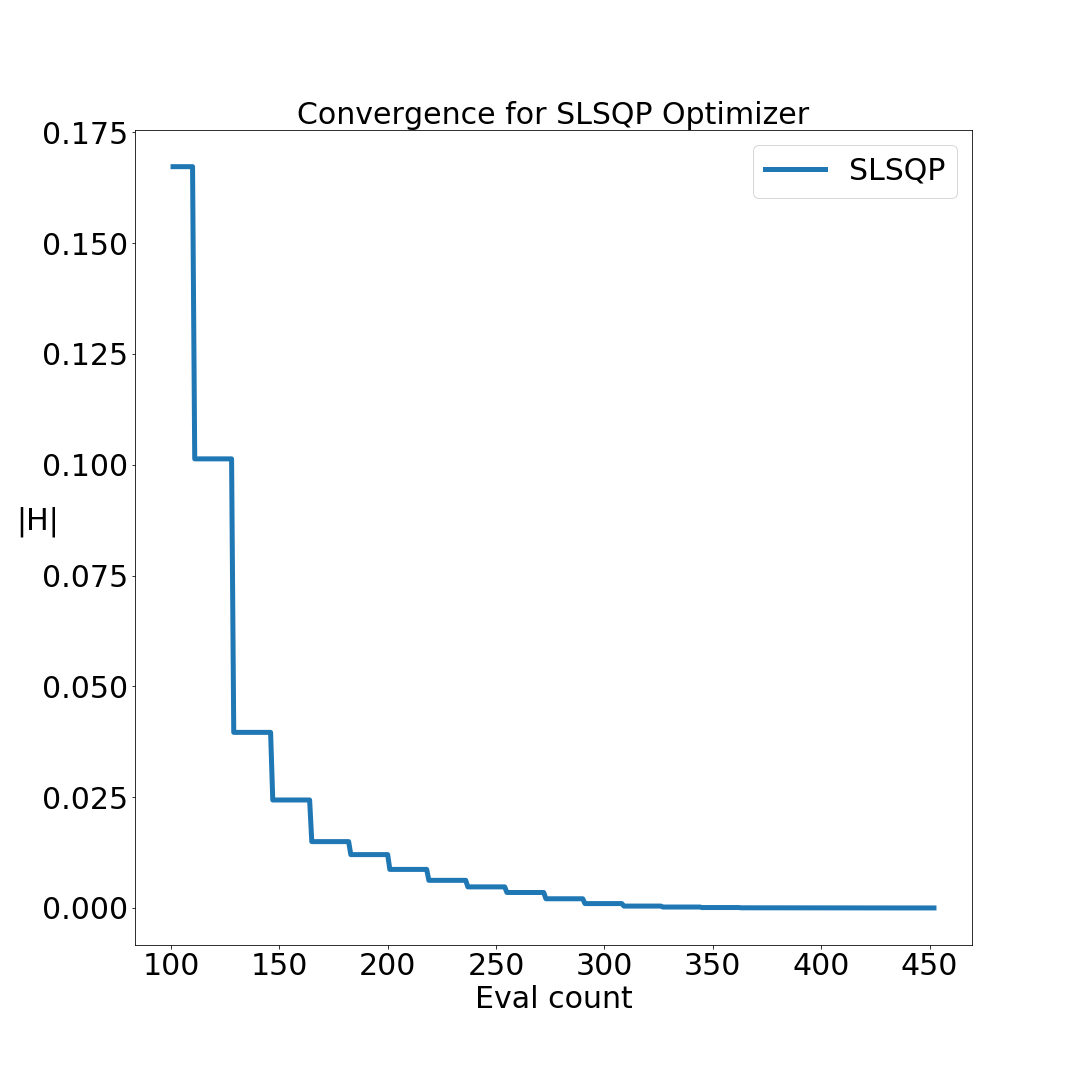}
\endminipage\hfill
\minipage{0.5\textwidth}
  \includegraphics[width=\linewidth]{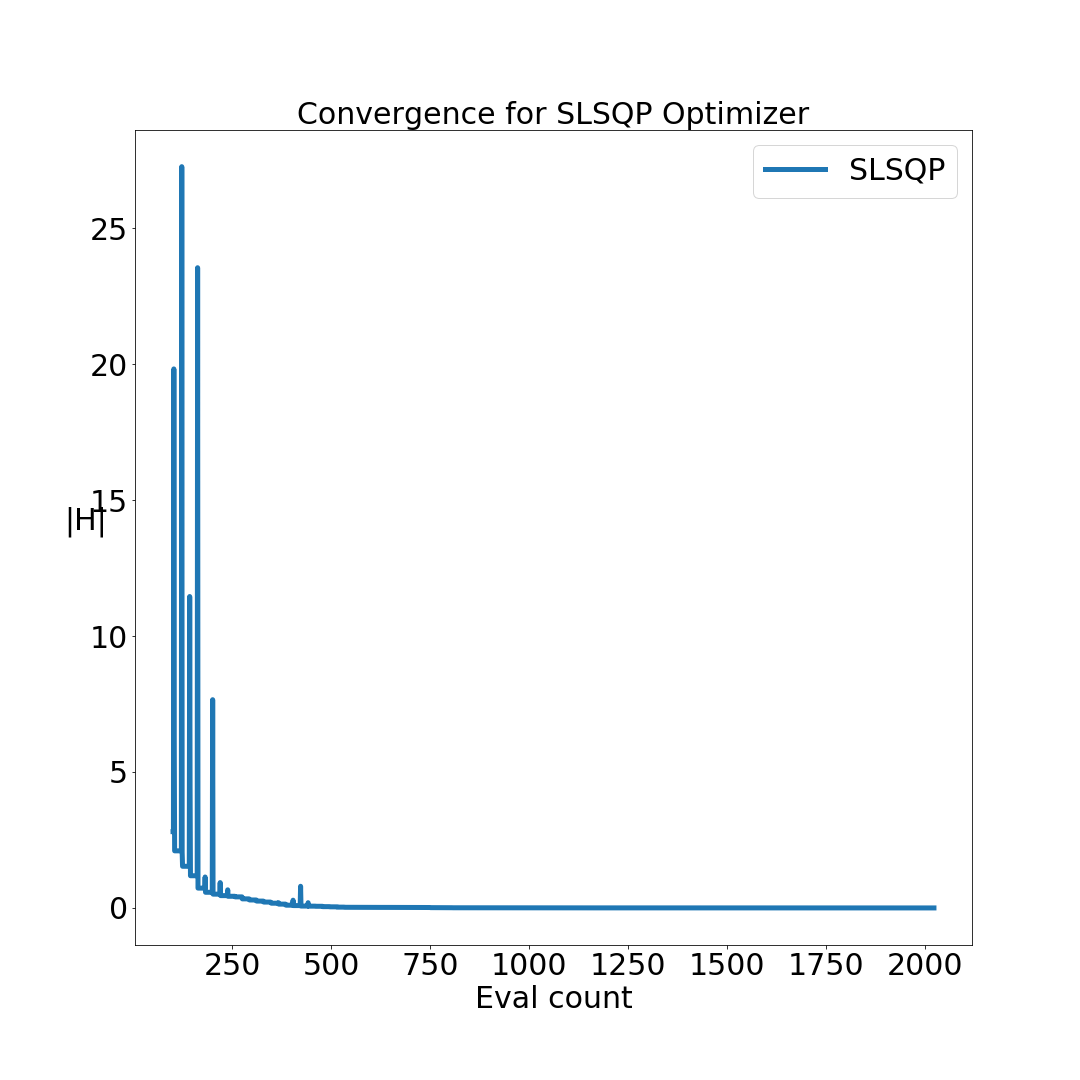}
\endminipage\hfill
\caption{Convergence graph for the quantum VQE computation of the magnitude of the Hamiltonian constraint for the 2D Charged String black hole with (left) $Q=1$ using the position basis, (right)  $Q=2$. using the the oscillator basis. 
}
\end{figure}

\begin{table}[ht!]
\centering
\begin{tabular}{|l|l|l|l|l|l|l|}
\hline
Charged String & basis      & Qubits  &  Paulis  & Exact & Discrete & VQE Result \\ \hline
$\abs{[H,M]}$ with $Q=1$ & pos & 4 & 72  & $0.0$ &  $0.0$ & $7.47395170\times 10^{-7}$ \\ \hline
$\abs{[H,M]}$ with $Q=2$ & osc  & 4 & 72  & $0.0$ &  $0.0$ & $0.00031898$ \\ \hline

\end{tabular}
\caption{\label{tab:BasisCompare}  VQE results for the lowest eigenvalue of the absolute value of the commutator of the Hamiltonian and Mass  operators for the 2D charged string  black hole using the oscillator basis. The Hamiltonian was mapped to 4-qubit operators for $(w,z)$ coordinates in mini-superspace. The quantum circuit for each simulation utilized an \(R_y\) variational form, with a fully entangled circuit of depth 3. The backend used was a state-vector simulator. The VQE results were obtained using the state-vector simulator with no noise and the Sequential Least Squares Programming (SLSQP) optimizer.}
\end{table}

\begin{figure}[htb!]
\centering
\minipage{0.5\textwidth}
  \includegraphics[width=\linewidth]{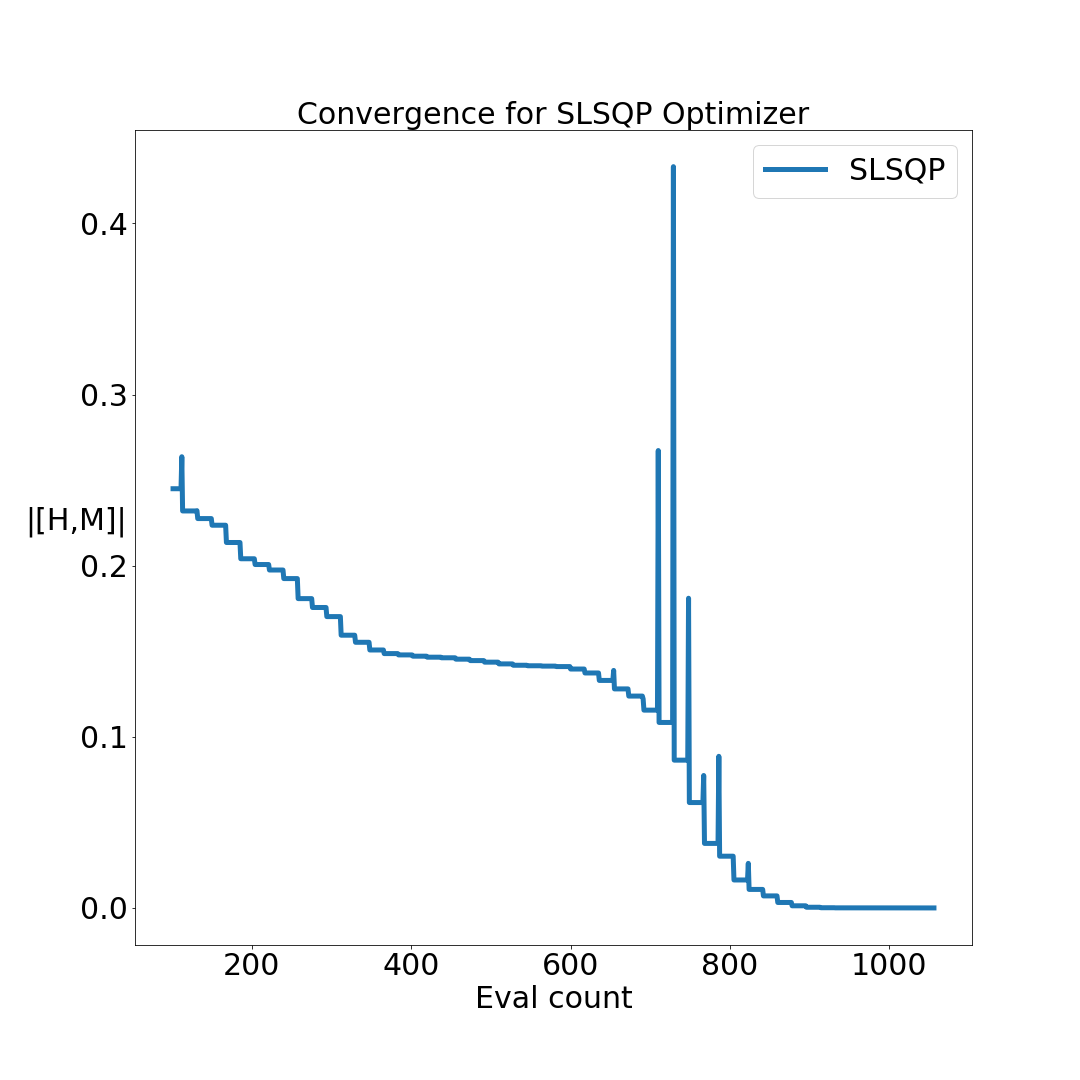}
\endminipage\hfill
\minipage{0.5\textwidth}
  \includegraphics[width=\linewidth]{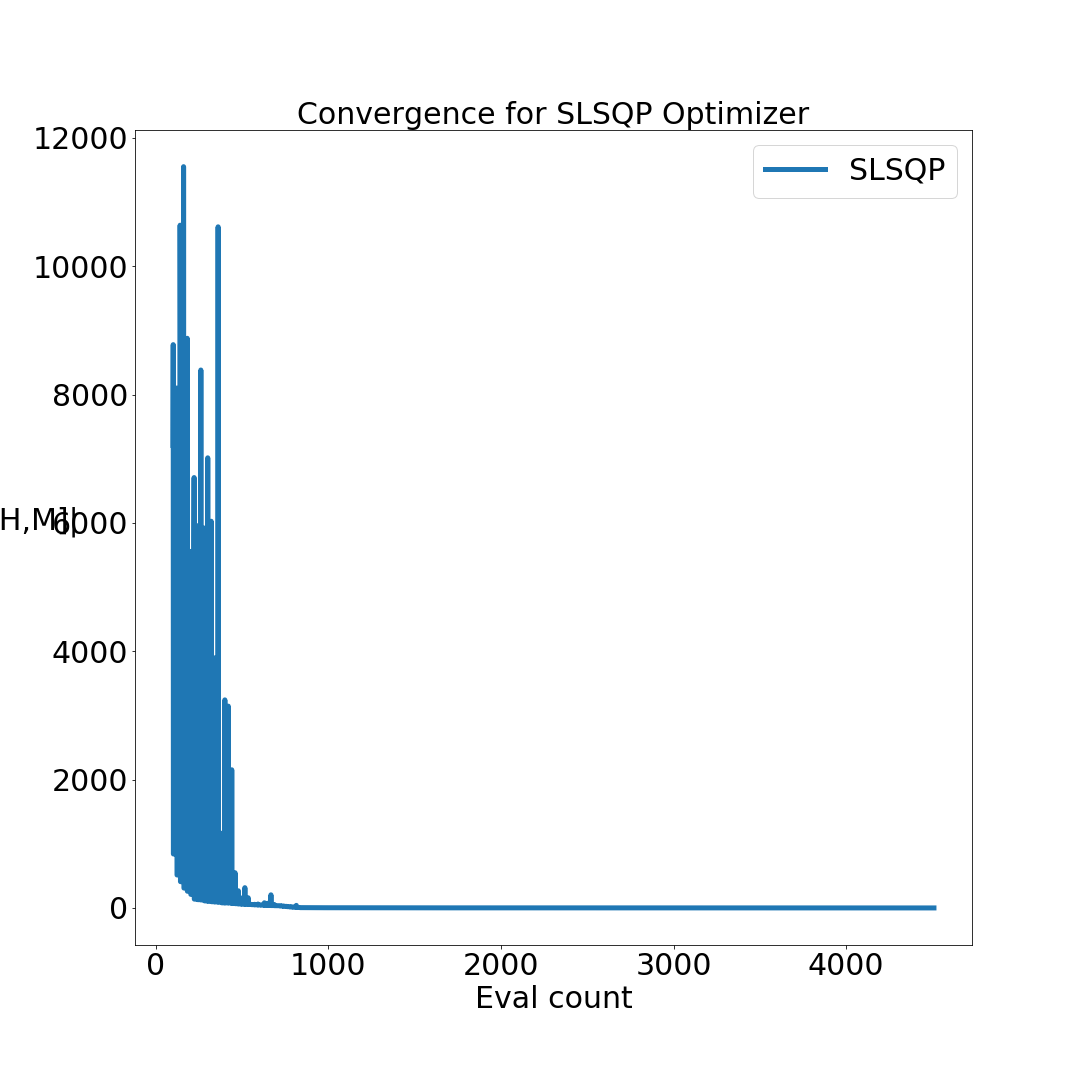}
\endminipage\hfill
\caption{Convergence graph for the quantum VQE computation of the magnitude of the commutator of the Hamiltonian constraint with the Mass operator for the 2D Charged String black hole with (left) $Q=1$ using the position basis, (right)  $Q=2$. using the the oscillator basis. 
}
\end{figure}

\section{Conclusions}

We applied the VQE quantum algorithm to find the lowest mass quantum eigenstate state and mass eigenvalue  of four types of black holes: the 3D rotating BTZ black hole, 4D charged RN black hole, 4D charged RN-dS black hole and the 2D charged string black hole. In all cases we were able to obtain accurate results using these methods. In the future we plan to apply extensions of the VQE method to find the quantum states of higher mass eigenstates and compare with WKB wave functions that capture behavior of classical black holes. We would also like to increase the number of qubits of the computations in order to consider the midi-superspace computations of the Hamiltonian and mass operators, as well as to include matter fields such scalar and fermionic fields, and treat the more complicated 4D Kerr and 4D Kerr-dS black holes. Matter fields with two types of gauge groups one visible and one hidden are especially interesting in studying the behavior of black holes in the presence of matter and dark matter in the early Universe with both matter and dark matter treated fully quantum mechanically.
\newpage

\section*{Acknowledgements}
We would like to to thank Amy Joseph, Mohammad Hassan and Enrico Rinaldi for useful discussions. Michael McGuigan would like to thank Satomi Ogawa for support and inspiration during the completion of this project.

\end{document}